\documentclass[aip,apl,reprint]{revtex4-1}

\usepackage{graphicx}
\usepackage{dcolumn}
\usepackage{bm}

\usepackage[utf8]{inputenc}
\usepackage[T1]{fontenc}
\usepackage{mathptmx}
\usepackage{etoolbox}
\usepackage{amsmath}
\usepackage{braket}
\usepackage{physics}
\makeatletter
\def\@email#1#2{%
 \endgroup
 \patchcmd{\titleblock@produce}
  {\frontmatter@RRAPformat}
  {\frontmatter@RRAPformat{\produce@RRAP{*#1\href{mailto:#2}{#2}}}\frontmatter@RRAPformat}
  {}{}
}%
\makeatother
\begin{document}

\preprint{AIP/123-QED}

\title{Strain-Engineered Deterministic Quantum Dots for Telecom O-Band Emission Using Buried Stressors}

\author{I. Limame}
\email{imad.limame@tu-berlin.de}
\affiliation{Institute of Physics and Astronomy, Technical University of Berlin, Hardenbergstraße 36, D-10623 Berlin, Germany}

\author{C.-W. Shih}
\author{K. Gaur}
\author{M. Podhorsk\'{y}}
\author{S. Tripathi}
\author{S. Wijitpatima}
\author{A. Koulas-Simos}
\author{C. C. Palekar}
\affiliation{Institute of Physics and Astronomy, Technical University of Berlin, Hardenbergstraße 36, D-10623 Berlin, Germany}

\author{P. Klenovsk\'{y}}
\affiliation{Department of Condensed Matter Physics, Masaryk University, Czech Republic}
\affiliation{Czech Metrology Institute, Okru\v{z}n\'{i} 31, 63800 Brno, Czech Republic}

\author{S. Reitzenstein}
\email{stephan.reitzenstein@physik.tu-berlin.de}
\affiliation{Institute of Physics and Astronomy, Technical University of Berlin, Hardenbergstraße 36, D-10623 Berlin, Germany}

\date{\today}

\begin{abstract}
The deterministic realization of quantum light sources operating at telecom wavelengths is essential for long-distance fiber-based quantum communication and distributed quantum computing. In this work, we demonstrate that telecom O-band emission can be achieved from site-controlled InGaAs/GaAs quantum dots (QDs). Our concept utilizes a buried AlAs/Al$_2$O$_3$ stressor layer with the unique feature that induces a well-defined and controllable tensile strain field at the growth surface, enabling both a redshift of QD emission to the $\sim$1.3~\textmu m range and site-selective nucleation at the mesa centers. This concept eliminates not only the need for strain-reducing layers (SRLs), which are known to degrade optical coherence, but also provides spatial control and spectral tunability. The grown telecom QDs show pure single-photon emission with $g^{(2)}(\tau) = (5.0 \pm 1.0) \times 10^{-2}$ at 4 K and $(2.8 \pm 0.3) \times 10^{-1}$ at 77~K, demonstrating the quantum nature and thermal stability of the emitters. The emission characteristics of complex excitonic states are analyzed using 8-band $k \cdot p$ and configuration-interaction modeling, which quantitatively reproduces the experimental observations. Finally, we present a theory-supported strategy to further redshift the emission toward the center of the O-band and beyond by employing a multi-buried-stressor approach. This combined framework of experiment and theory establishes the buried stressor concept as a scalable route toward highly coherent, position-controlled O-band quantum emitters compatible with industrial photonic integration.
\end{abstract}
\maketitle


\section{Introduction}

Quantum technologies use the principles of quantum mechanics to enable secure communication and advanced information processing with single photons as information carriers~\cite{Gisin2007,Long2007}. One promising approach in this field involves semiconductor quantum dots (QDs) as on-demand quantum light sources~\cite{Senellart2017, Heindel2023}. Acting as almost ideal two-level systems, these quantum emitters can produce single photons with high purity \cite{Sapienza2015,Somaschi2016,Schweickert2018, Hanschke2018}, high indistinguishability,~\cite{Ding2016, Hauser2025-hx, Margaria2025-tz}, and close-to-ideal entanglement fidelity \cite{Mller2014,Wang2019-vw, Lettner2021-ce} on demand. However, so far these best performance parameters have been limited to the 780 nm and 930 nm spectral range achieved with standard GaAs and InGaAs based QDs. On the other hand, for real-world applications in fiber-based quantum networks, single-photon sources (SPS) emitting in the telecom O-band and C-band with minimum dispersion and optical attenuation, respectively, are most important. In this context, a key advantage of self-assembled QDs, if compared for instance to other quantum emitters in solid state systems such as NV-centers in diamond and defect emitters in 2D materials, is their wavelength tunability, which is usually enabled by composition, size and strain control during epitaxial growth.

Driven by the high application potential, intensive research efforts have been directed to QDs with emission in the telecom O- and C-bands. Shifting the emission wavelength in this spectral range has been enabled by using complex growth concepts based on strain reducing layers (SRLs) and metamorphic buffer (MB) layers~\cite{Semenova2008, Olbrich2017}. Despite comprehensive growth optimizations, the photon indistinguishability of telecom O-band QDs is still limited to values below 20\% which can be attributed to spectral jitter induced by charged defect states at the SRL interfaces \cite{Kim2016, Srocka2020}. More recently, lattice-matched host materials such as InAlGaAs have been used to overcome this issue for C-band QDs and allowed to observe photon indistinguishability up to $(91 \pm 0.2)\%$ using longitudinal acoustic (LA) phonon assisted excitation \cite{Hauser2025-hx}. 

Another important issue in the field of QD-based quantum light sources is the scalable device integration of such emitters, which is usually hindered by their random positions in the growth plane resulting from the self-assembled nature of their fabrication. To overcome this limitation, it is essential to develop site-controlled QD (SCQD) growth, which has been mastered for QDs emitting in the usual 930 nm wavelength range \cite{Schneider2009-lb, Jons2013-ky, Grose2020-qa} and also at telecom wavelengths \cite{Haffouz2018-ql} based on nanohole and nanowire arrays. However, both nanohole- and nanowire-based approaches face inherent challenges, such as limited spectral tunability, surface-related charge noise, and difficulties in scalable device integration. In particular, nanohole QDs suffer from surface proximity, which leads to spectral diffusion and reduced quantum efficiency \cite{Albert2010}, while nanowire emitters often exhibit large structural inhomogeneity and incompatibility with standard epitaxial processing \cite{Mntynen2019}. In contrast, the buried stressor approach enables the nucleation of high-quality SCQDs on an unetched epitaxial surface, providing simultaneous spatial and spectral control through strain engineering without introducing surface defects \cite{Gaur2025-gw}. Succeeding in growing high quality SCQDs also at telecom wavelengths will enable the scalable fabrication and integration of QDs into nanophotonic devices like circular Bragg grating (CBG) resonators for emission enhancement and photonic integrated quantum circuits for applications in quantum communication and photonic quantum computing as recently shown in the 930 nm spectral range \cite{Gaur2025}.

In this work, we demonstrate the epitaxial growth of site-controlled InGaAs/GaAs QDs exhibiting pure single-photon emission in the telecom O-band without the need of a SRL. For this purpose we use the buried stressor concept as an advanced and versatile epitaxial method which uniquely combines the position-controlled nucleation of QDs with a redshift of their wavelength to the O-band range of 1.3 \textmu m.    
In fact, although it is known that buried stressors induce strain modulation at the growth surface, facilitating localized nucleation of QDs with adjustable density~\cite{Strittmatter2012,Limame2024b}, we go an important step beyond existing research by demonstrating that the tensile strain imposed by the stressor not only leads to position control but also shifts the QD emission to the telecom O-band. The buried stressor method therefore overcomes the critical limitations of the established methods of SRL and MB growth for telecom QDs. This method enables control over spatial positioning, emitter density, and emission wavelength to be achieved simultaneously \cite{Strittmatter2012, Kaganskiy2019, Limame2024b}. 

The technological and experimental work reported here is complemented by employing an advanced computational framework integrating~\textbf{\textbf{\textbf{\textbf{k}}$\cdot$\textbf{p}}} theory and the configuration interaction method~\cite{Millington-Hotze2025}. This approach enables us to obtain a profound understanding of the strain-controlled emission properties of the QDs through modelling their optical characteristics. It enables accurate predictions of the QD composition and size, providing deeper insights into the structural properties via calculations of the excitonic fine structure splitting (FSS) and binding energies of charged excitons $X^{\pm}$ and biexcitons $XX$ with respect to the neutral exciton $X^{0}$. As an outlook, we introduce a theory-guided multi–buried-stressor approach that enables further redshifting of the emission wavelength in future experiments, while simultaneously providing enhanced and tunable control over the surface strain profile. Our work establishes a foundation for scalable quantum light sources operating in the telecom O-band, leveraging SCQDs for precise emission control. These sources hold significant promise for integration into fiber-based quantum networks, enabling high-performance, fiber-based quantum communication with deterministic and reproducible device architectures.

\section{Materials and Methods}

The fabrication of the site-controlled telecom O-band QDs starts with a metal-organic chemical vapor deposition (MOCVD) growth of a template structure: a 300\,nm GaAs buffer, ten $\lambda/4n$ Al\textsubscript{0.9}Ga\textsubscript{0.1}As/GaAs distributed Bragg reflector (DBR) mirror pairs (110\,/\,95\,nm) on a 400\,\textmu m n-doped GaAs wafer, designed to enhance the photon extraction efficiency (PEE). Subsequently, a 50\,nm grading layer, a 30\,nm AlAs layer serving as the buried stressor after lateral oxidation, and an inverted grading layer capped with 80\,nm GaAs are grown. Next, square mesas with lateral dimensions between 20.4 \textmu m and 21.4 \textmu m, arranged with a pitch of 254 \textmu m and 67 nm size increment, are patterned via UV-lithography and inductively coupled plasma reactive-ion etching (ICP-RIE). Subsequently, selective wet oxidation at 420 °C is used to form oxide apertures acting as buried stressors with well-defined size. After oxidation, the structure is overgrown with a 50\,nm thick GaAs layer, a 0.35\,nm InGaAs layer, and a final 100\,nm GaAs cap after a 90\,s growth interruption. Crucially, this epitaxial layer design does not include an SRL or MB layers, both of which are essential ingredients of conventional growth schemes for telecom wavelength QDs. As such, our approach crucially simplifies the growth of telecom QDs and avoids SRL or MB layer-induced defects, which are known to lead to spectral jitter and degrade the quantum optical properties of such quantum emitters \cite{Wyborski2023-rm}. Notably, in comparison with Ref.~\cite{Limame2024b}, where a maximum emission wavelength of 1150~nm was achieved, the InGaAs QD seed layer employed in the present work features a substantially lower indium content (50\% versus 63\%), a reduced thickness (1.0~nm instead of 1.2~nm), and a significantly extended growth interruption time (90~s compared to 40~s). While the reduced indium concentration and thinner InGaAs wetting layer (WL) would, in isolation, be expected to induce a blueshift of the QD emission, the strong tensile strain introduced by the buried-stressor, together with the more than doubled growth interruption time, not only compensates this effect but results in an overall redshift exceeding 110~nm (from 1150~nm to 1260~nm). These growth parameters were deliberately optimized to shift the emission toward the telecom O-band while simultaneously reducing the areal density of QDs. Further details of the epitaxial growth procedure and parameter optimization are provided in the Supplementary Information (SI), Section 1.

\section{Results}

The optical properties of the grown SCQDs were studied by cathodoluminescence (CL) and micro-photoluminescence (\textmu PL) spectroscopy at 20 and 4 K, respectively. Additionally, single-photon emission was investigated using photon autocorrelation experiments.

	\begin{figure*}[t]  
		\centering
		\includegraphics[width=\textwidth]{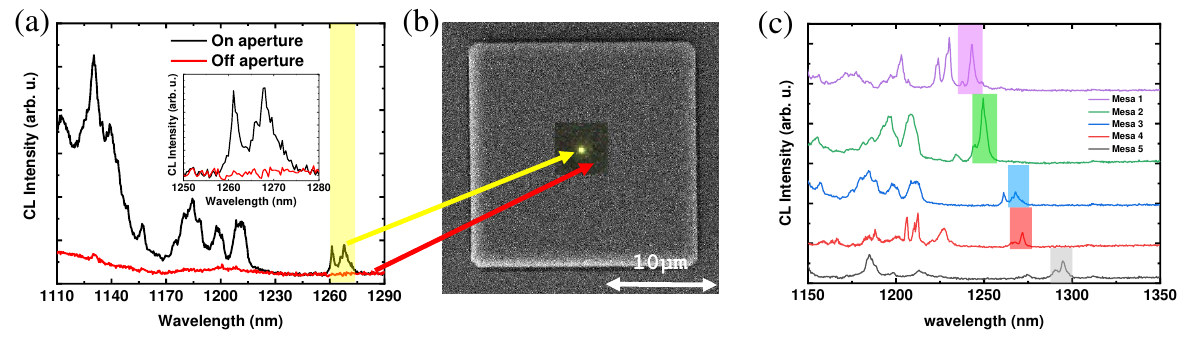}  
		\caption{Optical characterization of O-band SCQDs using CL. (a) CL spectrum of the mesa's central region (black trace) and from an off-center region (red trace). The inset shows a magnified view of the QD emission in the telecom O-band. The spectral segment highlighted in yellow, spanning from 1260 to 1280 nm, indicates the emission from SCQDs, as seen in the CL intensity map in (b). (b) CL map superimposed on a SEM image of the mesa. (c) Waterfall spectra showing the emission from the center region of five different mesas numbered from 1 to 5 with nominal mesa sizes of 21.07~\textmu m, 21.00~\textmu m, 21.20~\textmu m, 21.34~\textmu m, and 21.07~\textmu m, respectively.}
		\label{Fig:CL}
	\end{figure*}

Figure \ref{Fig:CL}~(a)~and~(b) display CL spectra and a combined scanning electron microscopy (SEM) image and CL intensity map of a selected SCQD mesa structure. As indicated by the arrows, the black spectrum corresponds to the central region of the mesa, while the red spectrum represents the off-center region. The yellow-highlighted section of the spectra (shown in the inset), spanning 1260–1280 nm (which is the lower end of the telecom O-band), corresponds to emissions from the most redshifted SCQDs directly above the aperture. Emission at shorter wavelengths (1100–1220 nm) is likewise observed at the mesa center, as shown in Fig.~\ref{Fig:CL}(a). This emission also originates from SCQDs, but most likely correspond to dots with reduced indium incorporation. This behavior indicates highly site-selective growth of SCQDs emitting in the O-band and lower wavelengths in the center of the mesa. The selective growth of QDs is attributed to the combination of a reduced InGaAs WL thickness, extended growth interruption time, and the strong buried-stressor-induced surface tensile strain compared to unstrained GaAs of up to 0.4\% above the aperture, all of which contribute to significant indium accumulation at the center of the mesa. Noteworthy, the InGaAs WL, with a thickness below the critical threshold for QD formation (approximately 4 monolayers or 1.17 nm) at 50\% indium content~\cite{Ustinov2003}, prevents the formation of O-band QDs on the non-strained GaAs regions of the mesa, while promoting nucleation above the strained aperture. The CL intensity map, overlaid on the SEM image in Fig.~\ref{Fig:CL}~(b), clearly shows a higher concentration of QD emission in the O-band within the yellow-highlighted spectral region of 1260 to 1280 nm at the mesa center. This confirms the localization and strain-mediated control of QD nucleation. Notably, the spatial and spectral control enabled by this approach demonstrates its high efficiency in tuning the emission properties of SCQDs, eliminating the need for an additional SRL or MB layer. The absence of QD emission from the off-center region (see red CL spectrum in Fig.~\ref{Fig:CL}~(a)) within the 1110–1290 nm wavelength range underscores the suitability of the buried stressor method for deterministic SPS fabrication and laser applications at telecom wavelengths.

As the deterministic positioning of the SCQDs is a key property of the buried stressor approach, we have investigated 21 structures with localized emission above 1200\,nm, as described in detail in the SI (Fig. S3). The spatial offsets of the emitters relative to the mesa center were extracted from CL maps. The resulting distributions yield mean displacements of $\bar{x}=-(25\pm47)\,\mathrm{nm}$ and $\bar{y}=-(130\pm63)\,\mathrm{nm}$ with standard deviations of $\sigma_x=217\,\mathrm{nm}$ and $\sigma_y=288\,\mathrm{nm}$. These values indicate that the SCQDs are well centered within the mesa structures with typical radial offsets on the order of $\sim300\,\mathrm{nm}$, confirming the high positioning accuracy of the buried stressor approach. However, these values remain larger than those reported in recent work on buried stressor SCQDs emitting at 930 nm \cite{Podhorsk2026}, where lateral displacements as small as $17^{+19}_{-17}\,\mathrm{nm}$ were demonstrated. We attribute the larger lateral offsets observed here to the comparatively large aperture size, which results in a non–single-peak tensile strain distribution.
    
Interestingly, the proposed buried stressor concept provides another appealing feature of emission control: as demonstrated in Figure~\ref{Fig:CL}(c), the emission wavelength of O-band SCQDs can be tuned by adjusting the mesa size, covering a range from 1250 nm to 1295 nm in the present case. Corresponding CL maps, which prove the site-selective growth, are presented in the SI. The semi-transparent boxes, color-coded to match each spectrum in panel (c), indicate the wavelength ranges used for the CL intensity maps. All maps reveal well-controlled positioning of the SCQDs, while also demonstrating the ability to spectrally tune the emission by more than 50 nm simply via the aperture size. This strain-controlled spectral tuning is primarily attributed to aperture-size-dependent tensile strain at the growth surface.

	\begin{figure*}[t]  
		\centering
		\includegraphics[width=\textwidth]{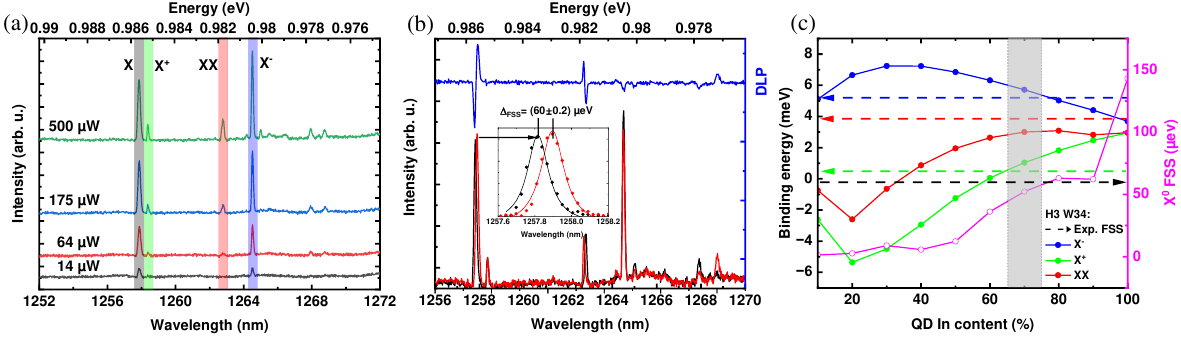}  
		\caption{(a) Waterfall plot displaying the \textmu PL emission from a selected SCQD under 1130~nm pulsed excitation, with pump powers ranging from 14 to 500~\textmu W at 4~K. Insets show Gaussian fits of the exciton ($X$) and negatively charged exciton ($X^-$) lines. (b) Polarization-resolved $\mu$PL spectra of the investigated SCQD at a pump power of 500~$\mu$W for polarization angles of 0$^\circ$ (black) and 90$^\circ$ (red). The inset shows a zoom of the exciton line, revealing an FSS of (60.0 $\pm$ 0.2)~$\mu$eV. The blue curve represents the degree of linear polarization (DLP). (c) The left axis shows CI-calculated binding energies of $X^-$, $X^+$, and $XX$ relative to $X^{0}$ (set to 0~meV) as a function of indium content for a QD with 3~nm height and 34~nm width. The gray box marks the $\sim$70\% indium range with the best agreement to experiment. The right axis (cyan) displays the simulated FSS of $X^{0}$ for the same geometry. The measured FSS is indicated by the dashed line and arrow, showing excellent agreement for 3~nm height, 34~nm base width, and 70\% indium (gray region).}
		\label{Fig:muPL}
	\end{figure*}

In the following, we present and discuss the optical properties of a single selected InGaAs/GaAs SCQD emitting in the telecom O-band. The spectroscopic corresponding data of an SCQD located in the center of a 21.3 \textmu m mesa is shown in Fig.~\ref{Fig:muPL}, where panel (a) displays a waterfall plot of the \textmu PL spectra of the investigated emitter under non-resonant excitation (1130 nm laser wavelength) with powers ranging from 14 to 500 \textmu W at a temperature of 4~K. At an intermediate excitation power of 175~\textmu W, the single SCQD exhibits a prominent excitonic ($X$) emission line at 0.9858~eV (approximately 1257.8~nm) and a negatively charged trion ($X^-$) line at 0.9805~eV (1264.5~nm), with respective linewidths of (71.0 $\pm$ 1.8) \textmu eV and (92.0 $\pm$ 2.4) \textmu eV. Additionally, the biexcitonic ($XX$) emission line at 0.9818~eV (1262.8~nm) and the positively charged trion ($X^+$) at 0.9853~eV (1258.4~nm) have linewidths of (57.0 $\pm$ 4.5)~\textmu eV and (73.0 $\pm$ 6.4)~\textmu eV, respectively. We refer to the SI, Section 2 for details on identifying the excitonic nature of the lines. The observed emission linewidths are significantly narrower than those reported for non-position-controlled O-band InGaAs QDs grown via the SRL approach, which typically range between $70$ and $150\,\mu\mathrm{eV}$~\cite{Paul2015-jz, Olbrich2017, Holewa2020}. Those linewidths could potentially be reduced through resonant excitation schemes~\cite{Nawrath2019}. The neutral exciton ($X$) exhibits a linear power dependence ($\sim P^{0.91}$) and an FSS of (60.0 ± 0.2) \textmu eV, as determined from polarization-resolved measurements shown in the inset of Fig.~\ref{Fig:muPL}~(b). This FSS is consistent with previously reported values for similar InGaAs/GaAs QDs grown on GaAs substrates~\cite{Holewa2020}.

We have performed correlated multi-particle calculations to support the experimental studies and their interpretation using a combination of the eight-band ${\bf k}\cdot{\bf p}$ method~\cite{Bahder1990,Stier1999,Birner2007} and the configuration interaction (CI) calculations~\cite{Bryant1987,Troparevsky2008,Schliwa:09,Klenovsky2017,Millington-Hotze2025,Klenovsk2026}. 
In the calculations of single-particle states, first the simulation model structure is defined on a rectangular grid. The elastic strain is then solved in the entire simulation space using the continuum elasticity method~\cite{Birner2007}, followed by a self-consistent solution of the Poisson and eight-band ${\bf k}\cdot{\bf p}$ Schr\"{o}dinger equations that includes the effects of nonlinear piezoelectricity~\cite{Bester:06,Beya-Wakata2011,Klenovsky2018}.

As a result, the single-particle eigenenergies $\mathcal{E}_k^{(e)}$ and $\mathcal{E}_l^{(h)}$ of electrons and holes, respectively, as well as the corresponding eigenfunctions $\psi_k^{(e)}$ and $\psi_l^{(h)}$ are obtained, with $k$ and $l$ labeling the states. From the set $\{\psi_k^{(e)},\psi_l^{(h)}\}$ the Slater determinants (SDs) labeled as $\left|D_m^{\rm M}\right>$, where $M\in\{{\rm X}, {\rm X}^-, {\rm X}^+, {\rm XX}\}$, are assembled. Next, multi-particle states are formed as $\Psi_i^{\rm M}(\mathbf{r}) = \sum_{\mathit m=1}^{n_{\rm SD}} \mathit \eta_{i,m} \left|D_m^{\rm M}\right>$, where $n_{\rm SD}$ is the number of SDs considered and $\eta_{i,m}$ is the $i$-th CI coefficient that is found along with the CI eigenenergy $E_i^{\rm{M}}$ 
by solving the Schr\"{o}dinger equation $\hat{H}^{\rm{M}} \Psi_i^{\rm M}(\mathbf{r}) = E_i^{\rm{M}} \Psi_i^{\rm M}(\mathbf{r})$. The CI Hamiltonian reads $\hat{H}^{\rm{M}}_{mn}=\delta_{mn}\left(\mathcal{E}_m^{{\rm M}(e)}-\mathcal{E}_m^{{\rm M}(h)}\right)+\left<D_m^{\rm M}\right| \hat{V}^{\rm{M}} \left|D_n^{\rm M}\right>$, where $\delta_{mn}$ is the Kronecker delta and $\mathcal{E}_m^{{\rm M}(e)}$ [$\mathcal{E}_m^{{\rm M}(h)}$] stands for sum of all single-particle electron [hole] eigenvalues corresponding to eigenstates contained in $\left|D_n^{\rm M}\right>$ for complex $M$.
Furthermore, $\left<D_m^{\rm M}\right| \hat{V}^{\rm{M}} \left|D_n^{\rm M}\right>=\mathcal{N}\sum_{ijkl}V^{\rm{M}}_{ij,kl}$ for $\{i,j\}\in S_m$ and $\{k,l\}\in S_n$. The sets $S_m$ and $S_n$ contain indices of single-particle wavefunctions in SDs $\left|D_m^{\rm M}\right>$ and $\left|D_n^{\rm M}\right>$, respectively. The normalization factor is $\mathcal{N}=1/4$ except if $m=n$ and $\{i,j\}=\{k,l\}$ in which case $\mathcal{N}=1/2$. Furthermore, $V^{\rm{M}}_{ij,kl}$ reads 
\begin{equation}
\label{eq:CIintegral}
V^{\rm{M}}_{ij,kl}\equiv(1-\delta_{ij})(1-\delta_{kl})\,q_iq_j\left(J^{\rm M}_{ij,kl} - K^{\rm M}_{ij,lk}\right),
\end{equation}
where $\delta_{ij}$ and $\delta_{kl}$ are the Kronecker deltas, thus, the first two brackets ensure that each single-particle state in SDs occurs only once; $q_i,q_j\in\{-1,1\}$, mark the signs of the charges represented by densities $|\psi_i|^2$ and $|\psi_j|^2$. The parameters $J^{\rm M}_{ij,kl}$ and $K^{\rm M}_{ij,lk}$ in Eq.~\eqref{eq:CIintegral} are the direct and the exchange Coulomb integrals between $\left<\psi_i\psi_j\right|$ and $\left|\psi_k\psi_l\right>$. During the CI computation the rate $\Gamma^{\rm M}_{i}$ of the radiative transition of state $i$ and the corresponding lifetime are evaluated employing the Hellmann-Feynman theorem~\cite{Dirac1927,Stier1999,Andrzejewski2010,Gaweczyk2017}.

        \begin{figure*}[t]
			\centering
			\includegraphics[width=\textwidth]{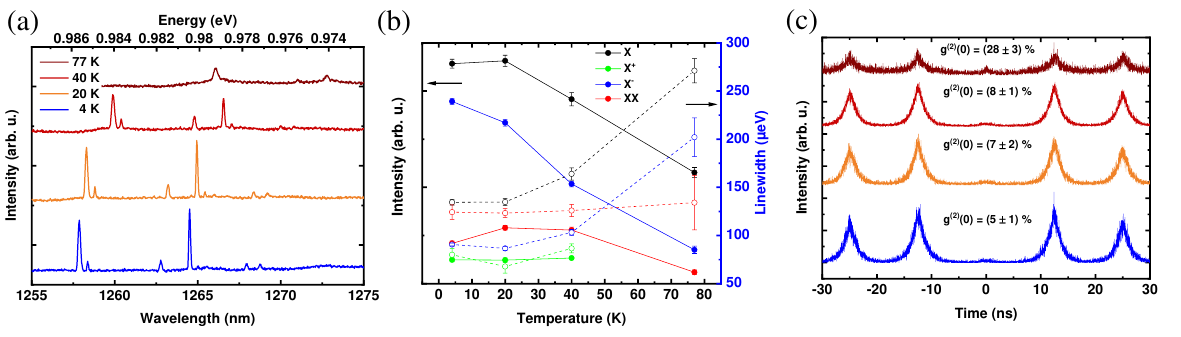}  
			\caption{(a) Waterfall plot showing \textmu{}PL spectra of the investigated SCQD under pulsed laser excitation at 1130 nm and a constant pump power of 140 \textmu{}W, measured at four different temperatures of 4 K, 20 K, 40 K, and 77~K. (b) \textmu{}PL intensity (solid circles) and linewidth (open circles) for the four identified excitonic transitions — ($X$, black), ($X^+$, green), ($XX$, blue), ($X^-$, red)—as a function of temperature. (c) Second-order photon autocorrelation function for the investigated SCQD line at a pump power of 140~\textmu W. We observe clean single-photon emission up to 40~K, with a $g^{(2)}(\tau)$ value of $(8.0 \pm 1) \times 10^{-2}$. At 77~K, the $g^{(2)}(\tau)$ value increases to $(2.8 \pm 0.3) \times 10^{-1}$.}
			\label{Fig:TempSer}
		\end{figure*}

The calculations of the binding energies of $X^-$, $X^+$, and $XX$ relative to $X^{0}$ as a function of indium content in QD with height of 3~nm and basis length of 34~nm are given in Fig.~\ref{Fig:muPL}~(c). Those calculations were performed for a CI basis of twelve electron and twenty four hole single-particle states~\cite{Birner2007} using the singles-doubles CI approximation (SDCI)~\cite{Schliwa:09}. The latter approximation was employed to reduce the numerical burden of the correlated CI calculations and produce almost the same results as that for the unapproximated CI~\cite{Schliwa:09}. 

In Fig.~\ref{Fig:muPL}~(c) we plot the calculated binding energies and the FSS as function of indium content. Our calculations reproduce very well the measured binding energies of the $X^-$, $X^+$, and $XX$ (the experimental values are marked by colored dashed arrows) relative to $X^{0}$ (set to 0~meV) for an indium content of 70~\% as indicted in the gray area. Similarly, we observe excellent agreement between experiment and theory for the FSS which matches again in the grayish area. Note that the FSS calculations considered the multipole expansion of the exchange interaction~\cite{Takagahara2000,Krapek2015}. We obtain the best agreement between calculated and measured FSS for the 3~nm high QD with basis length of 34~nm for a QD composed of 70\% of indium content. This value is consistent with our expectations, as the InGaAs QD seed layer contains 50\% indium. In combination with the extended growth interruption time, this promotes enhanced indium incorporation. Together with the tensile strain field, this leads to an increased effective indium content in the SCQDs.

Next, we examine the temperature-dependent optical and quantum optical properties of the selected SCQD, focusing on its prominent excitonic ($X$) transition. The related optical measurements were conducted at 4, 20, 40, and 77~K.  The latter two temperatures were chosen due to their relevance for practical SPS applications. While portable Stirling cryostats can readily achieve 40~K, making this temperature interesting for stand-alone SPSs~\cite{Schlehahn2018,Musial2020,Gao2022}, 77~K corresponds to the boiling point of liquid nitrogen and offers a cost-effective cooling alternative. 

Figure~\ref{Fig:TempSer}~(a) illustrates the temperature-dependent \textmu PL spectra of the SCQD under pulsed excitation at 1130~nm with a constant pump power of approximately 140~\textmu W. Due to the temperature dependence of the band gap, a clear redshift of the emission peak (the exciton, $X$) is observed as the temperature increases from 4 to 77~K. As shown in Fig.~\ref{Fig:TempSer}~(b), the intensity of all emission lines diminishes with increasing temperature (solid circles), primarily due to thermal escape of the carriers from the QD potential~\cite{Dusanowski2014}. Furthermore, emission linewidths broaden significantly at higher temperatures, as indicated by the open circles in Fig.~\ref{Fig:TempSer}~(b). This broadening is mainly attributed to increased phonon interactions, which not only scatter carriers within the QD but also induce dephasing of the excitonic states \cite{Stock2011-ad, Denning2020}.

The single-photon purity, quantified via the second-order autocorrelation function g$^{(2)}$(0), was evaluated for the excitonic emission over the studied temperature range, as shown in Fig.~ \ref{Fig:TempSer} (c). At 4~K, the QD shows a g$^{(2)}$(0) value of $(5 \pm 1) \times 10^{-2}$, indicating a single-photon purity 1-g$^{(2)}$(0) = 95\%. These values are slightly lower than those reported for InGaAs QDs grown via the SRL method at comparable temperature ranges~\cite{Holewa2020,Srocka2020}.  However, compared to Purcell-enhanced emitters integrated into CBG structures, the single-photon purity reported here is higher~\cite{Kolatschek2021-zz}.

	\begin{figure*}[t]
			\centering
			\includegraphics[width=\textwidth]{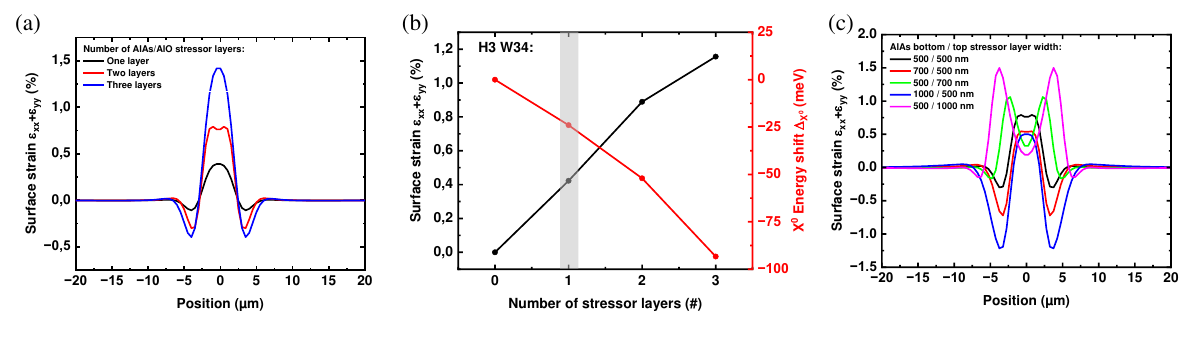}  
			\caption{(a) Calculated strain profiles at the growth surface for one (black), two (red), and three AlAs/Al$_2$O$_3$ stressor layers for an AlAs aperture of 500 nm. (b) Left: in-plane surface strain, $\varepsilon_{xx} + \varepsilon_{yy}$; right: corresponding strain-induced energy shift of a nominal QD (height: 3~nm, width: 34~nm, indium content: 70\%), shown as a function of the number of AlAs/Al$_2$O$_3$ stressor layers (one, two, and three).  (c) Calculated strain profiles for a two-layer AlAs/Al$_2$O$_3$ stressor configuration, illustrating the effect of varying the lower and upper aperture widths. Curves correspond to the following aperture combinations (lower/upper): 500/500\,nm (black), 700/500\,nm (red), 500/700\,nm (green), 1000/500\,nm (blue), and 500/1000\,nm (magenta), highlighting how aperture geometry tunes the local tensile strain.}
			\label{Fig:Strain}
		\end{figure*}

As the temperature increases, the single-photon purity decreases, reaching 72\% at 77~K. That decline in single-photon purity is attributed to multiple factors. First, higher temperatures enhance phonon interactions within the material, leading to phonon-induced carrier capture and release at defect states in the surrounding semiconductor matrix, which contribute to increased multi-photon events~\cite{Ortner2005,Dusanowski2014,Olbrich2017}. Second, elevated temperatures enable carriers to escape from the QD confinement and become captured and released by nearby QDs or defect states, introducing uncorrelated background emissions~\cite{Holewa2020}. Furthermore, at elevated temperatures, phonon sidebands contribute more significantly to the emission spectrum which leads to a broadening of the emission lines, as seen in Fig.~\ref{Fig:TempSer}~(b). That spectral broadening complicates the isolation of single-photon emission from background or multi-photon contributions, increasing the probability of detecting multiple photons simultaneously within one excitation pulse. These combined effects lead to a progressive decrease in single-photon purity as the temperature rises, consistent with observations reported for other QD systems~\cite{Olbrich2017,Holewa2020}.

As an outlook to future technological optimizations, we address the emission wavelength of the SCQDs via additional numerical modelling. For most of the experimentally studied emitters, the emission wavelength reaches only the lower bound of the telecom O-band (starting at 1260~nm). We attribute this limitation primarily to two factors: (i) the compressive strain imposed by the GaAs host matrix and (ii) the non-optimized stressor design with compressive strain below 0.5\% employed in the present study. Regarding the first aspect, embedding of QDs in a low-indium-content surrounding layer, i.e. an SRL, has been widely used and studied for non-positioned InGaAs QDs as a route to partially relieve compressive strain to reach the O-band spectral range under tensile surface strain \cite{Paul2017-ef, Olbrich2017}. However, this approach has been shown to adversely affect the optical quality of the emitters, including linewidth broadening and reduced single-photon purity \cite{Srocka2020}. As an alternative, AlInGaAs has recently emerged as a promising host material, demonstrating high optical quality at telecom C-band wavelengths while maintaining performance comparable to weakly strained InGaAs QDs on GaAs emitting near 900~nm~\cite{Hauser2025-hx}. Regarding the second aspect, to achieve emission deep in the O-band in combination with high scalability of QD growth, we propose the implementation of a multi-buried-stressor design to enhance the tensile surface strain and thereby promote increased indium accumulation at the mesa center during growth. This approach enables a local strain environment with significantly higher tensile strain, which is essential for pushing the emission wavelength deeper into the telecom O-band and beyond. Figure~\ref{Fig:Strain}~(a) shows the calculated surface strain profiles, $\varepsilon_{xx} + \varepsilon_{yy}$, obtained using continuum elasticity theory~\cite{Gaur2025-gw}, for structures incorporating one, two, and three AlAs/Al$_2$O$_3$ stressor layers with a fixed aperture diameter of 500~nm. The schematics of the corresponding layer stacks and a detailed description of the simulation methodology are provided in the SI. 

The black curve in Fig.~\ref{Fig:Strain}~(a) corresponds to the experimentally investigated single-stressor structure and exhibits a single, well-defined strain maximum with a tensile strain magnitude of approximately 0.39--0.40~\% relative to unstrained GaAs. Similar simulation results have been reported previously, along with their experimental validation by Raman spectroscopy~\cite{Limame2024b}. Introducing a second 30~nm AlAs stressor layer separated by a 10~nm AlGaAs spacer leads to a twofold increase in tensile strain in the center of the mesa, reaching approximately 0.79~\%, as shown by the red curve. 

Notably, the strain profile evolves from a single-peak to a double-peak configuration. Although such a double-peak strain distribution is undesirable for deterministic positioning of SCQDs within photonic nanostructures~\cite{Gaur2025}, it introduces an additional degree of freedom in strain engineering. By tailoring the Ga concentration (0--5\%) in the Al(Ga)As layers, the oxidation rate and thus the effective aperture size can be controlled independently for each stressor layer~\cite{Choquette1997}. This enables fine-tuning of both the magnitude and spatial profile of the surface strain $\varepsilon_{xx} + \varepsilon_{yy}$, offering enhanced design flexibility for application-specific quantum photonic devices, as shown in Fig.~\ref{Fig:Strain}(c) and discussed below. We identify this tunability as a key advantage of the multi-buried-stressor approach for scalable and deterministic quantum emitter platforms.

Remarkably, the addition of a third stressor layer not only increases the tensile strain magnitude to approximately 1.41~\%, more than a threefold enhancement compared to the single-stressor case, but also restores a single, mono-peak strain profile. This strain configuration is highly desirable for precise control over SCQD density and positioning. Moreover, the substantially increased tensile strain enables SCQD formation using InGaAs wetting layer thicknesses below the critical thickness for QD nucleation on unstrained GaAs, thereby significantly improving growth selectivity. This aspect is particularly critical for the realization of SCQD-based lasers, where non-positioned QDs located outside the optical mode center introduce excess optical losses due to absorption~\cite{Kaganskiy2019, Shih2024}.

The dependence of the maximum tensile surface strain on the number of stressor layers is summarized in Fig.~\ref{Fig:Strain}~(b). The corresponding calculated strain values were applied to a nominal SCQD with a height of 3~nm, a width of 34~nm, and an indium concentration of 70~\%, yielding the calculated emission redshifts shown in Fig.~\ref{Fig:Strain}~(b) (red symbols). Starting from an unstrained QD with those structural parameters, the introduction of a single AlAs/Al$_2$O$_3$ stressor layer (gray shaded region), as realized experimentally in this work, results in an expected redshift of approximately 24~meV (31~nm). This estimate accounts solely for the externally applied strain and does not include additional redshifting due to indium accumulation during growth. Nevertheless, the latter effect is partially captured by the elevated indium concentration (70\%) assumed for the SCQD, compared to the nominal growth concentration of 50\%. Upon adding a second and third stressor layer, the calculated emission redshifts increase to approximately 52~meV (67~nm) and 93~meV (120~nm), respectively. Given the experimentally observed emission around 1260~nm at 4~K, these results indicate that a double-stressor design would enable emission near the center of the telecom O-band under the current growth conditions, while a three-stressor architecture would allow access to the long-wavelength edge of the O-band. These findings establish multi-stressor engineering as a powerful and scalable route toward the buried stressor growth of SCQDs operating across the full telecom O-band and beyond.

As discussed above, the Gallium composition in the Al(Ga)As stressor layer provides a powerful and technologically accessible tuning parameter to control the effective stressor width, and thus the magnitude and spatial profile of the induced strain field. Figure~\ref{Fig:Strain}(c) systematically illustrates this capability for a double-stressor structure by varying the lower and upper aperture widths. The investigated geometries (lower/upper) are 500/500\,nm (black), 700/500\,nm (red), 500/700\,nm (green), 1000/500\,nm (blue), and 500/1000\,nm (magenta).

The symmetric 500/500\,nm configuration (black) serves as a reference and reproduces the strain profile shown in Fig.~\ref{Fig:Strain}(a) (red curve). Increasing the width of the lower stressor layer to 700\,nm (red) and 1000\,nm (blue) results in a pronounced enhancement of the compressive strain from $-0.28\%$ to $-0.72\%$ and $-1.22\%$, respectively. At the same time, the tensile strain at the mesa center decreases from $0.76\%$ to $0.54\%$ and $0.50\%$, accompanied by a transition from a weak double-peak to a single-peak profile. This clearly demonstrates that the lower stressor layer predominantly governs the compressive strain landscape while moderating and spatially smoothing the tensile maximum.

In contrast, increasing the width of the upper stressor layer (700\,nm, green; 1000\,nm, magenta), while keeping the lower layer fixed at 500\,nm, leads to a fundamentally different behavior. Here, the tensile strain is strongly enhanced and reshaped, evolving from a weakly double-peaked profile with a maximum of $0.79\%$ to a pronounced double-peak distribution reaching $1.06\%$ (700\,nm) and up to $1.50\%$ (1000\,nm). This enhancement is accompanied by a lateral broadening of both tensile and compressive regions, while the compressive strain slightly decreases. These results highlight that the upper stressor layer primarily controls the localization, symmetry, and magnitude of the tensile strain maximum.

Overall, the multi-stressor approach introduces a highly versatile design space for engineering strain fields with tailored amplitude and spatial characteristics. This level of control is particularly relevant for SCQDs, where the precise positioning, confinement potential, and emission wavelength are directly governed by the local strain environment. In this context, the ability to independently tune compressive and tensile strain components enables deterministic control over nucleation sites, improved spatial accuracy, and enhanced uniformity of optical properties.

However, while the emission redshift achieved via the multi-buried-stressor approach follows a clear and predictable trend, its experimental implementation presents nontrivial challenges. In particular, replacing a single AlAs/Al$_2$O$_3$ stressor layer with multiple stressors of larger total thickness can be problematic. Such an approach may lead to mechanical instability of the upper layers due to excessive accumulated strain, resulting in partial delamination or peeling of the overgrown structure. These considerations highlight the importance of distributing the strain across multiple thinner stressor layers rather than concentrating it within a single thick layer, thus ensuring structural integrity and reproducible device fabrication.

\section{Conclusion}

In this work, we demonstrate the strong potential of the buried-stressor approach for realizing site-controlled quantum dots (SCQDs) emitting in the telecom O-band. Unlike conventional epitaxial schemes relying on strain-reducing or metamorphic buffer layers, this approach enables precise site-selective nucleation together with a strain-engineered redshift of the QD emission to $\sim$1.3~\textmu m. Its compatibility with established VCSEL fabrication processes provides a scalable and industry-relevant route toward integrated single-photon sources for quantum communication.The observed thermal stability and single-photon purity of up to 72\% at 77~K highlight the robustness of the emitters, while local control of QD density opens pathways to high-$\beta$ microlasers and other nanophotonic devices with tailored confinement. Furthermore, the multi--buried-stressor concept enables additional redshifting of the emission, extending its applicability across the near-infrared spectral range. Combined with the excellent agreement between experiment and advanced modeling, these results establish the buried-stressor approach as a powerful platform for scalable, high-quality quantum emitters in the telecom regime.

\vspace{1em}

\textbf{Funding.} The authors gratefully acknowledge the financial support from the Volkswagen Foundation via the project NeuroQNet2, the German Research Foundation via INST 131/795-1 320 FUGG, from the Federal Ministry of Research, Technology and Space (BMFTR) via the project MultiCoreSPS (Grant No. 16KIS1819K), and from Berlin Quantum (BQ). P.K acknowledges funding from the European Innovation Council Pathfinder program under grant agreement No 101185617 (QCEED), funding by the Institutional Subsidy for the Long-Term Conceptual Development of a Research Organization granted to the Czech Metrology Institute by the Ministry of Industry and Trade of the Czech Republic and by the project Quantum Materials for applications in sustainable technologies, CZ.02.01.01/00/22\_008/0004572.

\vspace{1em}

\textbf{Acknowledgment.} The authors thank Kathrin Schatke, Praphat Sonka, Heike Oppermann, Stefan Bock, Lucas Rickert, and Martin von Helversen for their invaluable technical support and engaging scientific discussions, which greatly contributed to this research.

\vspace{1em}

\textbf{Disclosures.} The authors declare no conflicts of interest.

\vspace{1em}

\textbf{Data Availability Statement.} Data underlying the results presented in this paper are not publicly available at this time, but can be obtained from the authors upon reasonable request.

\vspace{1em}

\textbf{Supplemental document.} See Supplementary Information for supporting content.

\clearpage
\onecolumngrid

\section*{Supplementary Information: Strain-Engineered Deterministic Quantum Dots for Telecom O-Band Emission Using Buried Stressors}

I. Limame*, C.-W. Shih, K. Gaur, M. Podhorsk\'{y}, S. Tripathi, S. Wijitpatima, A. Koulas-Simos, C. C. Palekar, P. Klenovsk\'{y}, and S. Reitzenstein

\bigskip

The Supplementary Information (SI) provides additional details on the epitaxial growth and fabrication of the site-controlled quantum dots (SCQDs) in Section 1, as well as extended optical characterization, including cathodoluminescence (CL) and $\mu$PL data, in Section 2. Furthermore, it contains a comprehensive description of the theoretical model used to simulate the strain induced by the buried stressor and its impact on the excitonic complexes of the investigated SCQDs, along with additional simulation results.

\section{\label{sec:level1}Epitaxial growth and fabrication}
	
The fabrication of SCQDs emitting in the telecom O-band using the buried stressor approach involves a two-step metal-organic chemical vapor deposition (MOCVD) process and one nanofabrication step as described in the following.
	
\subsection{Template structure growth}
	
\begin{figure*}[h]  
    \renewcommand{\thefigure}{S1}
		\centering
		\includegraphics[width=\textwidth]{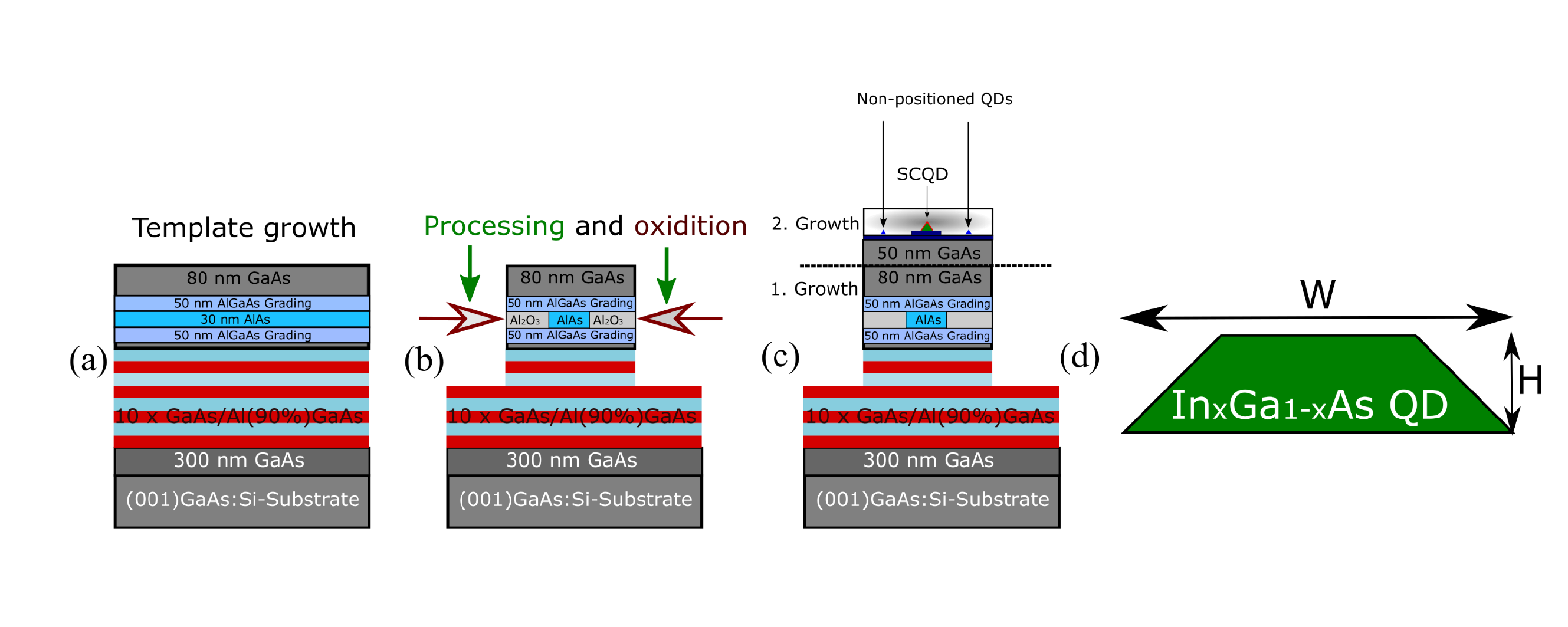}  
		\caption{(a) Epitaxial layer design consisting of a template structure that includes a 300 nm buffer layer, 10 pairs of GaAs/Al\textsubscript{0.9}Ga\textsubscript{0.1}As/GaAs DBR optimized for emission at 1300 nm, a 30 nm AlAs stressor layer, and an 80 nm GaAs cap. (b) After the UV lithography step, selective wet oxidation forms an unoxidized AlAs aperture at the center of the mesa. (c) The strain from this aperture facilitates the accumulation of indium atoms, leading to site-controlled nucleation of QDs above the aperture. Additionally, with a 90-second growth interruption, this strain redshifts the single QD emission. (d) The simulated structure of the truncated cone-shaped QD is shown, including its dimensions (height and width) and the indium concentration ($x$) within the dot.}
	\label{Fig:Structure}
	\end{figure*}

The process begins with the epitaxial growth of a template structure on a 400~nm thick, n-doped GaAs:Si (001) 2-inch wafer. A 300 nm GaAs buffer layer is deposited at 700~\textdegree C and 100~mbar, ensuring a high-quality epitaxial surface. Following this, a distributed Bragg reflector (DBR) composed of 10 pairs of $\lambda/4$ Al\textsubscript{0.9}Ga\textsubscript{0.1}As/GaAs layers, with nominal thicknesses of 110 nm and 95 nm, respectively, is grown to enhance photon extraction efficiency. Next, a 50 nm compositionally graded layer transitions from GaAs to 30 nm of AlAs, which later serves as the buried stressor. This is followed by an inverted grading layer and an 80 nm GaAs capping layer, which prevents surface oxidation (Fig. \ref{Fig:Structure}(a)).

\subsection{Nanofabrication}
	
The template is coated with AZ 701MIR photoresist (300~nm thick) via spin-coating, and ultraviolet lithography (UVL) is used to define square mesa patterns with side lengths of 20-21 \textmu m and a nominal 63~nm step size. The resist is post-baked at 110 \textdegree C, and the structures are developed using AZ 726 MIF. Inductively coupled plasma-reactive ion etching (ICP-RIE) transfers the pattern into the III-V semiconductor material, creating square mesas etched down to the fourth DBR mirror pair. The exposed AlAs layer is then selectively oxidized in a vacuum furnace at 420 \textdegree C and 50~mbar using a nitrogen and water vapor mixture. The oxidation process continues until the desired aperture size is achieved, resulting in a structure where a small, unoxidized AlAs aperture is surrounded laterally by Al\textsubscript{2}O\textsubscript{3} (Fig.~\ref{Fig:Structure}~(b)). The partial oxidation of the AlAs layer induces an in-plane strain due to the volume contraction associated with the formation of AlOx. This strain propagates up to the surface, resulting in a strain profile. The characteristics and effects of this strain field have been extensively studied, as detailed in~\cite{Limame2024b}. Moreover, this strain can be precisely monitored and optimized using Raman spectroscopy techniques. For instance, micro-Raman setup, as demonstrated in~\cite{Ries2024-cx, Ries2024-gp}, provides spatially resolved insights into the strain distribution across the mesa surface.
	
\subsection{SCQD growth}
After thorough cleaning, the second MOCVD step begins. A 50~nm GaAs layer is deposited at 700~\textdegree C, followed by a 0.3~nm In\textsubscript{0.5}Ga\textsubscript{0.5}As wetting layer (WL) grown at approximately 500~\textdegree C. A growth interruption of 90 seconds is applied before depositing a 1~nm GaAs layer. The temperature is then raised to 615~\textdegree C to desorb uncapped indium (In). A 100~nm GaAs barrier layer is subsequently deposited to minimize coupling between the SCQDs and surface states.  
	
The final structure is illustrated in Fig. \ref{Fig:Structure}(c). Figure~\ref{Fig:Structure}(d) shows the simulated structure of the truncated cone-shaped QD, including its dimensions. In the calculations, both the base width ($W$) and height ($H$) of the QD, as well as the indium content ($x$), are varied. The WL is omitted in all simulations presented here to isolate the effects of QD size and composition without introducing additional contributions from the WL.

\section{Optical results}

\begin{figure*}[h]  
    \renewcommand{\thefigure}{S2}
		\centering
		\includegraphics[width=\textwidth]{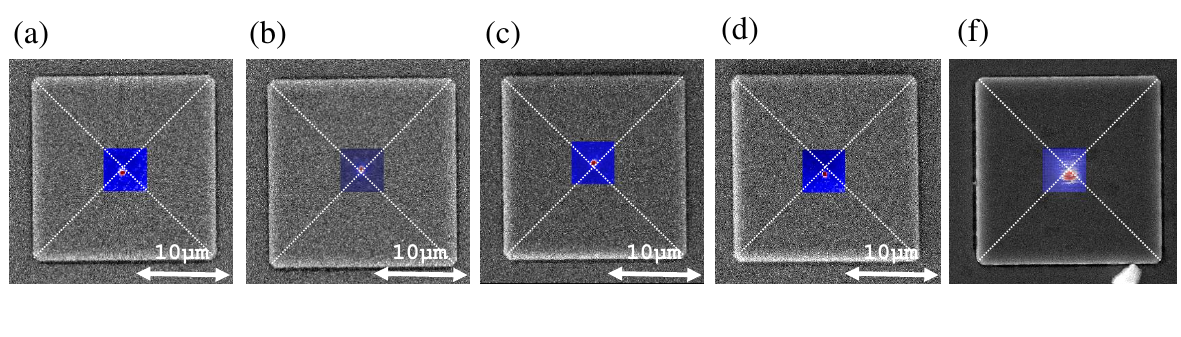}  
		\caption{(a–f) Combined SEM and CL maps of five different mesas  (Mesa 1-5 in Fig. 1 (c)) showing emission from SCQDs located at the center of each mesa near and within the telecom O-band. The white dashed lines indicate the mesa center, providing a visual reference that highlights the spatial offset of the SCQD emission.}
		\label{Fig:CL_Maps}
		
	\end{figure*}

\begin{figure*}[t]  
\renewcommand{\thefigure}{S3}
	\centering
	\includegraphics[width=\textwidth]{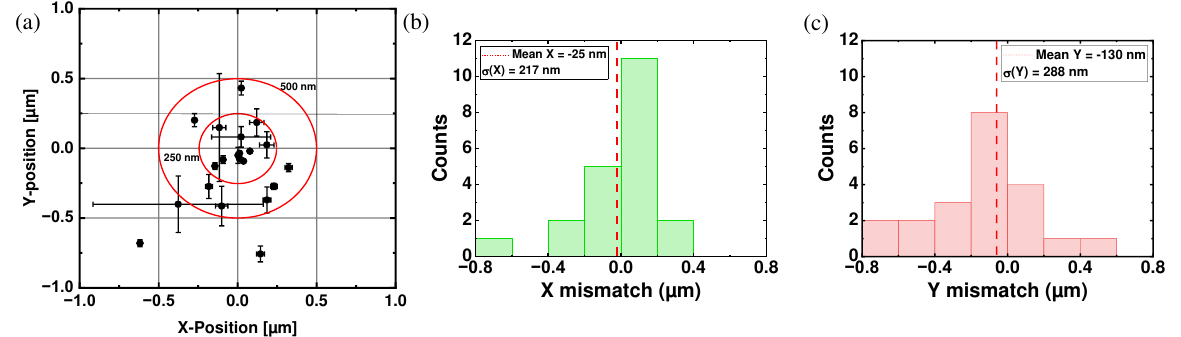}  
	\caption{(a) Spatial distribution of the extracted SCQD positions relative to the center of the respective mesa structures. The positions were determined from cathodoluminescence (CL) mapping. Two red reference circles with radii of $250\,\mathrm{nm}$ and $500\,\mathrm{nm}$ are included to visualize the spatial spread of the emitters around the mesa center. (b) Histogram of the $x$-coordinate displacement, evaluated with a bin width of $200\,\mathrm{nm}$. The distribution is centered at a mean displacement (red dashed line) of $x = -(25 \pm 47)\,\mathrm{nm}$ with a standard deviation of $217\,\mathrm{nm}$, indicating that the SCQDs are on average well aligned with the mesa center and show no pronounced systematic shift along the $x$-direction. (c) Histogram of the $y$-coordinate displacement, obtained using the same binning, yields a mean displacement (red dashed line) of $y = -(130 \pm 63)\,\mathrm{nm}$ and a standard deviation of $288\,\mathrm{nm}$, indicating a slightly broader spatial spread along the $y$-direction and a small systematic offset toward negative $y$ values.}
	\label{Fig:Position}

\end{figure*}

Figure \ref{Fig:CL_Maps} (a–f) present scanning electron microscopy (SEM) images and CL maps recorded from five different mesas (Mesa 1–5, as labeled in Fig. 1(c)) in the spectral window of 1240 to 1300 nm. Each map reveals a well-defined emission spot originating from a SCQD located precisely at the center of the mesa (see Fig.~\ref{Fig:Position}). The observed emission falls within the telecom O-band, confirming successful spectral tuning of the SCQDs to the target wavelength range. Among these structures, Mase 3 shown in panel (c) is the structure investigated in detail in the Results section of the main text.

As shown in Fig.~\ref{Fig:Position} (a) statistical evaluation was performed on the measured lateral displacements of 21 SCQDs relative to the mesa center $(0,0)$ for emission above 1200 nm. The distribution of positions yields mean displacements of $\bar{x}=-(25\pm47)\,\mathrm{nm}$ and $\bar{y}=-(130\pm63)\,\mathrm{nm}$ (mean $\pm$ standard error of the mean), indicating that the ensemble is centered close to the mesa origin with only a small systematic offset along the $y$-direction, as seen in Fig.~\ref{Fig:Position}(b) and (c), respectively. The standard deviations of the spatial distributions are $\sigma_x=217\,\mathrm{nm}$ and $\sigma_y=288\,\mathrm{nm}$, revealing a slightly broader spread along $y$ than along $x$. The median positions ($x_\mathrm{med}=12\,\mathrm{nm}$, $y_\mathrm{med}=-82\,\mathrm{nm}$) are close to the corresponding mean values, indicating no pronounced skewness or dominant outliers in the dataset. The mean radial displacement from the mesa center amounts to $\langle r \rangle = (297 \pm 236)\,\mathrm{nm}$ (standard deviation), corresponding to typical offsets below approximately $300\,\mathrm{nm}$. To further illustrate the spatial spread of the emitters, the plot in Fig.~\ref{Fig:Position} (a) includes two reference circles with radii of $250\,\mathrm{nm}$ and $500\,\mathrm{nm}$. These guides to the eye show that most SCQD positions lie within a $500\,\mathrm{nm}$ radius around the mesa center, while a considerable fraction is already contained within $250\,\mathrm{nm}$. A weak positive correlation between the $x$ and $y$ coordinates ($r=0.24$) indicates no pronounced directional alignment of the displacements. Overall, the statistical distribution confirms that the SCQDs are located close to the mesa center with moderate spatial spread and no significant anisotropy.

	\begin{figure*}[b]  
    \renewcommand{\thefigure}{S4}
		\centering
		\includegraphics[width=\textwidth]{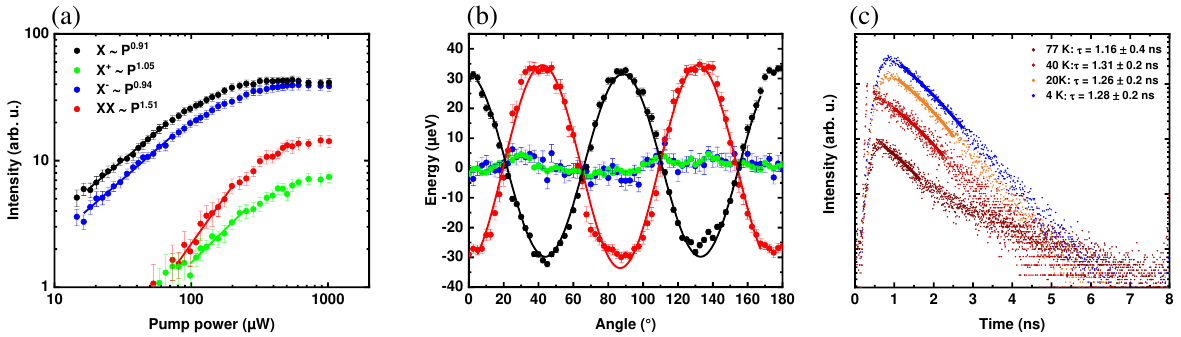}  
		\caption{(a) Double logarithmic plot illustrating the integrated \textmu PL intensity as a function of excitation power for the exciton (X, black data points), positively charged exciton ($X^{+}$, green data points), biexciton ($XX$, red data points), and negatively charged exciton ($X^{-}$, blue data points). Solid lines of corresponding colors indicate the respective slopes. (b) Polarization-resolved measurements of the aforementioned states reveal a fine structure splitting (FSS) of $\Delta_{\text{FSS}} = (60.0 \pm 0.2)~\mu\text{eV}$.}
		\label{Fig:Pow_Pol}
		
	\end{figure*}
	
To investigate the nature of the emission lines, power-dependent and polarization-resolved \textmu PL measurements were conducted, as shown in Fig.~\ref{Fig:Pow_Pol}~(a) and (b). These measurements allowed the identification of four distinct transitions associated with the studied SCQD: a neutral excitonic ($X$) state, a neutral biexcitonic ($XX$) state, and two charged excitonic states ($X^{+}$ and $X^{-}$). The neutral exciton (X), observed at 0.9858~eV (1257.8~nm), exhibits a nearly linear power dependence with an intensity scaling as $P \sim (0.91 \pm 0.02)$, as shown in Fig.~\ref{Fig:Pow_Pol}~(a). Additionally, the FSS of the $X$ and $XX$ states was determined to be $(60.0 \pm 0.2)$ \textmu eV, based on sinusoidal fits of the peak energy as a function of polarization angle, depicted in Fig.~\ref{Fig:Pow_Pol}~(b) (solid black and red lines). The biexciton ($XX$) emission, observed at 1262.8~nm, is distinguishable by its antiphase linear polarization relative to the exciton emission and its super-linear power dependence, following $P \sim (1.51 \pm 0.05)$, as shown in Fig.~\ref{Fig:Pow_Pol}~(a). This behavior is consistent with the expected two-photon cascade process from the biexciton to the exciton state. However, for the biexciton ($XX$), a quadratic power dependence ($I_{XX} \propto P^2$) is expected under ideal conditions, as two electron–hole pairs are required. The observed deviation from this behavior indicates non-ideal carrier capture dynamics, leading to a reduced scaling exponent. The charged excitonic states ($X^{+}$ and $X^{-}$), identified at 1264.5~nm and 1258.4~nm, respectively, show no measurable FSS in the polarization-resolved \textmu PL measurements and a linear power dependence of $P \sim (1.05 \pm 0.06)$ and ($0.94 \pm 0.02$), respectively. These results are consistent with the typical optical signatures of charged excitonic states in similar QD systems.

Time resolved \textmu PL measurements of excitonic emission at four distinct temperatures, 4, 20, 40, 77~K is shown in Fig.~\ref{Fig:Pow_Pol}~(c) and reveal a temperature independent decay time of $(1.24 \pm ~0.30)$ ~ns, which is typical for as-grown InGaAs/GaAs QDs \cite{Kolatschek2021-zz}. At 77~K, a biexponential decay profile emerges. This behavior is attributed to phonon interactions and increased background emission from temperature-activated defect states and the WL ~\cite{Xu2008,Braun2014}.
	
\section{Theoretical modelling}
    \begin{figure}[htbp]
    \centering
    \renewcommand{\thefigure}{S5}
        \includegraphics[width=85mm]{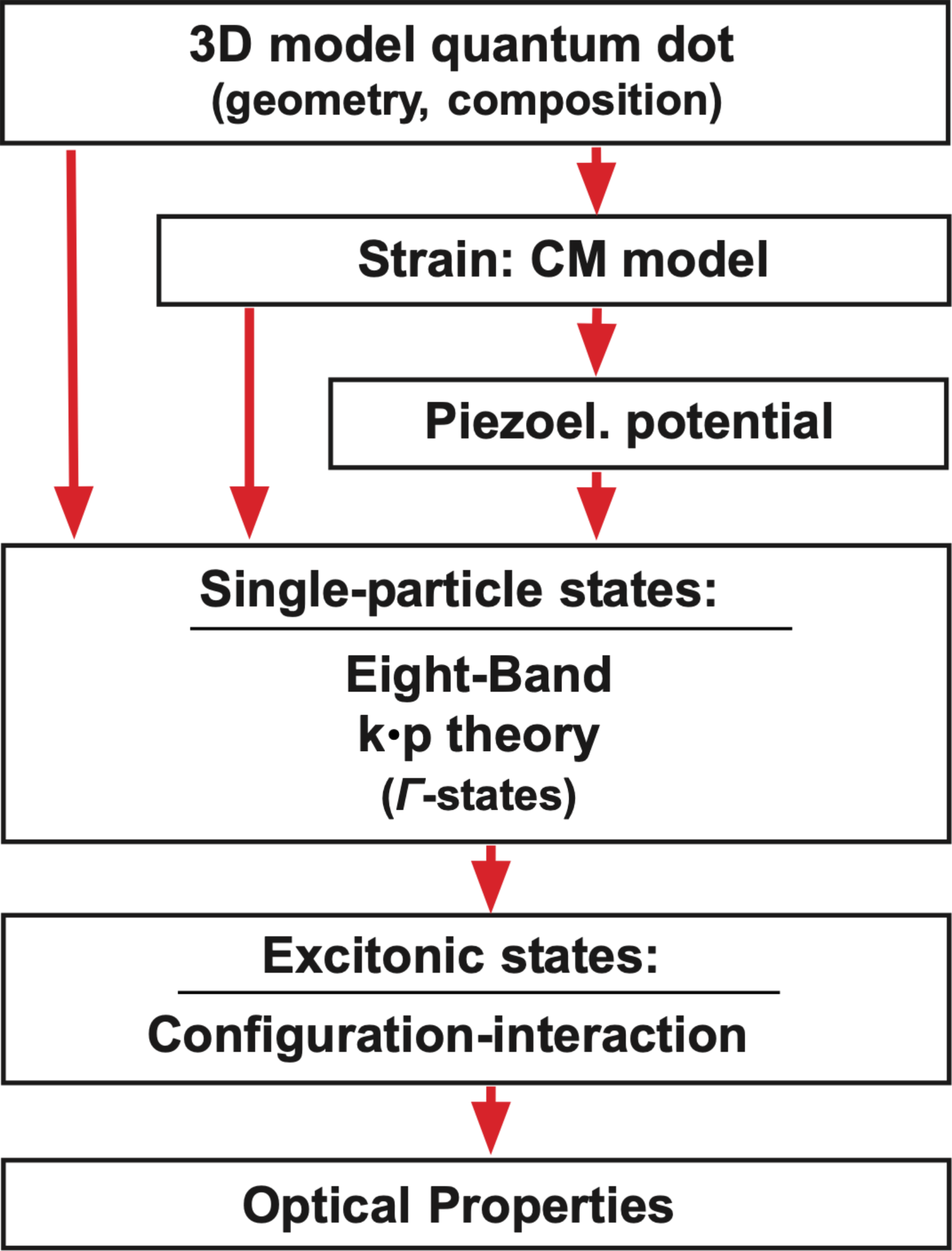}	
    	\caption{Method of the calculations in this work.}
    	\label{fig:theorMethod}
    \end{figure}

In this work, a combination of numerical methods is employed to model the structural, electronic, and optical properties of the SCQD system. The strain distribution at the QD growth surface is determined using the continuum elasticity model implemented in Nextnano++~\cite{Birner2007}, which accounts for lattice mismatch effects and the strain from the partially oxidized AlAs stressor layer. Additionally, the presence of a defect layer is incorporated by defining the layer within Nextnano++, where the Poisson equation is solved to calculate the resulting electrostatic potential. This step is essential for understanding the impact of charge trapping and internal fields on carrier confinement. To determine the single-particle states, the 8-band k·p envelope function approximation is applied, capturing the effects of strain, band mixing, and spin-orbit coupling. This method provides a realistic description of the QD energy levels and wavefunctions. Furthermore, the influence of Coulomb interactions on excitonic and multiparticle states is investigated using a Configuration Interaction (CI) approach, implemented in a custom-developed CI code. This method allows for the accurate modeling of electron-hole interactions, excitonic fine structure, and multi-exciton binding energies, enabling a detailed prediction of the emission spectra. By combining these computational techniques, we achieve a comprehensive and precise description of the structural, electronic, and optical properties of the QD system.

The whole method of computation of electronic states in the present work is sketched in Fig.~\ref{fig:theorMethod}. We now discuss the steps of the calculation in more detail.  

\subsection{Single-particle states}

We start by computing the single-particle states of the system. First, we implement the 3D QD model structure (size, shape, chemical composition). That is followed by the calculation of elastic strain by minimizing the total strain energy in the structure and subsequent evaluation of piezoelectricity up to non-linear terms.~\cite{Bester:06,Beya-Wakata2011,Klenovsky2018} The resulting strain and polarization fields then enter the eight-band $\mathbf{k}\!\cdot\!\mathbf{p}$ Hamiltonian.

In ${\bf k}\cdot{\bf p}$, implemented within the Nextnano++ computational suite~\cite{Birner2007}, we consider the single-particle states as linear combinations of $s$-orbital~like and $x$,~$y$,~$z$~$p$-orbital~like Bloch waves~\cite{Klenovsky2017,Birner2007} at $\Gamma$ point of the Brillouin zone,~i.e.,

    \begin{equation}
        \psi_{a_n}(\mathbf{r}) = \sum_{\nu\in\{s,x,y,z\}\otimes \{\uparrow,\downarrow\}} \chi_{a_n,\nu}(\mathbf{r})u^{\Gamma}_{\nu}\,,
    \end{equation}

where $u^{\Gamma}_{\nu}$ is the Bloch wavefunction of $s$- and $p$-like conduction and valence bands at $\Gamma$ point, respectively, $\uparrow$/$\downarrow$ marks the spin, and $\chi_{a_n,\nu}$ is~the~envelope function for $a_n \in \{ e_n, h_n \}$ [$e$ ($h$) refers to electron (hole)] of the $n$-th single-particle state.

Thereafter, the following envelope-function $\mathbf{k}\!\cdot\!\mathbf{p}$ Schr\"{o}dinger equation is solved

%
%
\begin{equation}
\label{eq:EAkp}
    \sum_{\nu\in\{s,x,y,z\}\otimes \{\uparrow,\downarrow\}}\left(\left[E_\nu^{\Gamma}-\frac{\hbar^2{\bf \nabla}^2}{2m_e}+V_{0}({\bf r})\right]\delta_{\nu'\nu}+\frac{{\hbar}{\bf \nabla}\cdot{\bf p}_{\nu'\nu}}{m_e}\right.
    \left. + \hat{H}^{\rm str}_{\nu'\nu}({\bf r})+\hat{H}^{\rm so}_{\nu'\nu}({\bf r})\right)\chi_{a_n,\nu'}({\bf r})=\mathcal{E}^{k\cdot p}_n\cdot \chi_{a_n,\nu'}({\bf r}),
\end{equation}

where the term in round brackets on the left side of the equation is the envelope function $\mathbf{k}\!\cdot\!\mathbf{p}$ Hamiltonian $\hat{H}_0^{k\cdot p}$, and $\mathcal{E}^{k\cdot p}_n$ on the right side is the $n$-th single-particle eigenenergy. Furthermore, $E_\nu^{\Gamma}$ is the energy of bulk $\Gamma$-point Bloch band $\nu$, $V_0({\bf r})$ is the scalar potential (e.g. due to piezoelectricity), $\hat{H}^{\rm str}_{\nu'\nu}({\bf r})$ is the Pikus-Bir Hamiltoninan introducing the effect of elastic strain,~\cite{Birner2007} and $\hat{H}^{\rm so}_{\nu'\nu}({\bf r})$ is the spin-orbit Hamiltonian.~\cite{Birner2007}
Further, $\hbar$ is the reduced Planck's constant, $m_e$ the free electron mass, $\delta$ the Kronecker delta, and $\nabla := \left( \frac{\partial}{\partial x}, \frac{\partial}{\partial y}, \frac{\partial}{\partial z} \right)^T$.

Furthermore, in the eight-band ${\bf k}\cdot{\bf p}$ model, the spin–orbit interaction is explicitly included through the coupling between conduction and valence bands. In particular, the valence band states are described within the total angular momentum basis $\ket{J, m_J}$ with $J = 3/2$ (heavy and light holes) and $J = 1/2$ (split-off band), where $m_J$ combines both spin and orbital angular momentum. As a result, the single-particle states $\psi_k^{(e)}$ and $\psi_l^{(h)}$ obtained from the ${\bf k}\cdot{\bf p}$ Hamiltonian represent mixed spin–orbital character. Consequently, spin is not a good quantum number in this basis and cannot be unambiguously separated or assigned to the single-particle orbitals used in subsequent configuration interaction (CI) calculations.

The aforementioned Schr\"{o}dinger equation is then solved self-consistently with the Poisson equation to improve the spatial position of electron and hole wavefunctions.~\cite{Birner2007} Note that the Poisson equation solver used in the single-particle calculations does not include Coulomb exchange.

\subsection{Configuration interaction}

The single-particle states computed by the aforementioned ${\bf k}\cdot{\bf p}$ are used as basis states for CI. In CI, we consider the multi-particle ($M$) states as linear combinations of the Slater determinants (SDs). In all generality, for any total number of fermions $N$ in the system, the $m$-th SD is written as:

\begin{equation}
\ket{D_m^M} = \frac{1}{\sqrt{N!}} \sum_{P} (-1)^P \, 
\psi_{P(i_1)}(\vb{r}_1) \, 
\psi_{P(i_2)}(\vb{r}_2) \, \cdots \, 
\psi_{P(i_N)}(\vb{r}_N),
\end{equation}

where $N \equiv N_e + N_h$, with $N_e$ ($N_h$) the number of electrons (holes) in the complex $M$ (e.g., $N_e = 2$, $N_h=1$ for the negative trion X$^-$). Moreover, $P$ marks the permutation of wavefunction indices.
SDs of the multi-particle states considered in this work are for the neutral exciton X

\begin{equation}
\label{eq:suppl:CIWavefunctionX}
\left|{\rm X}\right>=\sum^{n_e}_{i=1}\sum^{n_h}_{j=1}\eta^{\rm X}_{ij}
\begin{vmatrix}
\psi_{ei}(\vec{r}_{e}) & \psi_{ei}(\vec{r}_{h})\\
\psi_{hj}(\vec{r}_{e}) & \psi_{hj}(\vec{r}_{h})
\end{vmatrix},
\end{equation}
for positive trion $X^+$
\begin{equation}
\label{eq:suppl:CIWavefunctionX+}
\left|{\rm X}^+\right>=
\sum^{n_e}_{i=1}\,\sum^{n_h}_{\substack{j,k=1\\k>j}}\eta^{\rm X^+}_{ijk}
\begin{vmatrix}
\psi_{ei}(\vec{r}_{e})&\psi_{hj}(\vec{r}_{e})&\psi_{hk}(\vec{r}_{e})\\
\psi_{ei}(\vec{r}_{h1})&\psi_{hj}(\vec{r}_{h1})&\psi_{hk}(\vec{r}_{h1})\\
\psi_{ei}(\vec{r}_{h2})&\psi_{hj}(\vec{r}_{h2})&\psi_{hk}(\vec{r}_{h2})\\
\end{vmatrix},
\end{equation}
for negative trion X$^-$
\begin{equation}
\label{eq:suppl:CIWavefunctionX-}
\left|{\rm X}^-\right>=
\sum^{n_e}_{\substack{i,j=1\\j>i}}\,\sum^{n_h}_{k=1}\eta^{\rm X^-}_{ijk}
\begin{vmatrix}
\psi_{ei}(\vec{r}_{e1})&\psi_{ej}(\vec{r}_{e1})&\psi_{hk}(\vec{r}_{e1})\\
\psi_{ei}(\vec{r}_{e2})&\psi_{ej}(\vec{r}_{e2})&\psi_{hk}(\vec{r}_{e2})\\
\psi_{ei}(\vec{r}_{h})&\psi_{ej}(\vec{r}_{h})&\psi_{hk}(\vec{r}_{h})\\
\end{vmatrix},
\end{equation}
and for the neutral biexciton $XX$
\begin{equation}
\label{eq:suppl:CIWavefunctionXX}
\left|{\rm XX}\right>=
\sum^{n_e}_{\substack{i,j=1\\j>i}}\,\sum^{n_h}_{\substack{k,l=1\\k>l}}\eta^{\rm XX}_{ijkl}
\begin{vmatrix}
\psi_{ei}(\vec{r}_{e1})&\psi_{ej}(\vec{r}_{e1})&\psi_{hk}(\vec{r}_{e1})&\psi_{hl}(\vec{r}_{e1})\\
\psi_{ei}(\vec{r}_{e2})&\psi_{ej}(\vec{r}_{e2})&\psi_{hk}(\vec{r}_{e2})&\psi_{hl}(\vec{r}_{e2})\\
\psi_{ei}(\vec{r}_{h1})&\psi_{ej}(\vec{r}_{h1})&\psi_{hk}(\vec{r}_{h1})&\psi_{hl}(\vec{r}_{h1})\\
\psi_{ei}(\vec{r}_{h2})&\psi_{ej}(\vec{r}_{h2})&\psi_{hk}(\vec{r}_{h2})&\psi_{hl}(\vec{r}_{h2})\\
\end{vmatrix},
\end{equation}
where $n_e$ and $n_h$ mark the number of single-particle states for electrons and holes, respectively. Furthermore, the aforementioned SDs defined in the Eqs.~(\ref{eq:suppl:CIWavefunctionX}-\ref{eq:suppl:CIWavefunctionXX}) must be normalized, i.e. $\sum_m |\eta_{m}|^2=1$.

Using the aforementioned $\left|D_m^{\rm M}\right>$ the multi-particle trial wavefunction reads
\begin{equation}
    \Psi_i^{\rm M}(\mathbf{r}) = \sum_{\mathit m=1}^{n_{\rm SD}} \mathit \eta_{i,m} \left|D_m^{\rm M}\right>, \label{eq:CIwfSD}
\end{equation}
where $n_{\rm SD}$ is the number of Slater determinants $\left|D_m^{\rm M}\right>$, and $\eta_{i,m}$ is the $i$-th CI coefficient which is found along with the eigenenergy using the variational method by solving the Schr\"{o}dinger equation 
\begin{equation}
\label{CISchrEq}
\hat{H}^{\rm{M}} \Psi_i^{\rm M}(\mathbf{r}) = E_i^{\rm{M}} \Psi_i^{\rm M}(\mathbf{r}),
\end{equation}
where $E_i^{\rm{M}}$ is the $i$-th eigenenergy of the multi-particle state $\Psi_i^{\rm M}(\mathbf{r})$, and~$\hat{H}^{\rm{M}}$ is the CI Hamiltonian which reads
%
%
%
%
\begin{equation}
\label{eq:CIHamiltonian}
\hat{H}^{\rm{M}}_{mn}=\delta_{mn}\left(\mathcal{E}_m^{{\rm M}(e)}-\mathcal{E}_m^{{\rm M}(h)}\right)+\left<D_m^{\rm M}\right| \hat{V}^{\rm{M}} \left|D_n^{\rm M}\right>,
\end{equation}
where $\delta_{mn}$ is the Kronecker delta and $\mathcal{E}_m^{{\rm M}(e)}$ ($\mathcal{E}_m^{{\rm M}(h)}$) stands for sum of all single-particle electron (hole) eigenvalues corresponding to eigenstates contained in $\left|D_n^{\rm M}\right>$ for complex $M$. Furthermore, $\left<D_m^{\rm M}\right| \hat{V}^{\rm{M}} \left|D_n^{\rm M}\right>=\mathcal{N}\sum_{ijkl}V^{\rm{M}}_{ij,kl}$ for $i,j\in S_m$ and $k,l\in S_n$. The sets $S_m$ and $S_n$ contain indices of all single-particle wavefunctions in SDs $\left<D_m^{\rm M}\right|$ and $\left|D_n^{\rm M}\right>$, respectively. The parameter $\mathcal{N}$ is the normalization factor, see in the following. Furthermore, $V^{\rm{M}}_{ij,kl}$ is defined by
%
%
%
%
%
%
%
\begin{equation}
\label{eq:CoulombMatrElem}
\begin{split}
V^{\rm{M}}_{ij,kl}\equiv&(1-\delta_{ij})(1-\delta_{kl})\,q_iq_j\frac{e^2}{4\pi\epsilon_0}\iint\left(\frac{{\rm d}{\bf r}_1{\rm d}{\bf r}_2}{\epsilon_r(\mathbf{r}_1,\mathbf{r}_2)|{\bf r}_1-{\bf r}_2|}\right)\\
&\times\{\psi^*_i({\bf r}_1)\psi^*_j({\bf r}_2)\psi_k({\bf r}_1)\psi_l({\bf r}_2)
-\psi^*_i({\bf r}_1)\psi^*_j({\bf r}_2)\psi_l({\bf r}_1)\psi_k({\bf r}_2)\}\\
=&(1-\delta_{ij})(1-\delta_{kl})\,q_iq_j\left(J^{\rm M}_{ij,kl} - K^{\rm M}_{ij,lk}\right),
\end{split}
\end{equation}
where $\epsilon_0$ and $\epsilon_r(\mathbf{r}_1,\mathbf{r}_2)$ are the vacuum and spatially dependent relative permittivity, respectively, and $\delta_{ij}$ and $\delta_{kl}$ are the Kronecker deltas. Note that the terms in the first two brackets in Eq.~\eqref{eq:CoulombMatrElem} ensure that each single-particle state in SD occurs only once, thus preventing double counting. Furthermore, $q_i,q_j\in\{-1,1\}$ marks the sign of the charge of the quasiparticles in states with indices $i$ and $j$, respectively; $e$ is the elementary charge. The parameters $J^{\rm M}$ and $K^{\rm M}$ in Eq.~\eqref{eq:CoulombMatrElem} are direct and exchange Coulomb integrals. 
%
%

Now, since the single-particle states are orthonormal, one finds that in Eq.~\eqref{eq:CIHamiltonian} there are only three possible kinds of matrix elements in CI,~i.e.
\begin{equation}
  \begin{split}
\label{eq:CIHamiltonianSeparated}
\hat{H}^M_{nm} &\equiv \left<{D^M_m}\right|{\hat{H}^M}\left|{D^M_n}\right>  \\
&=   \begin{cases}
    \mathcal{E}_m^{{\rm M}(e)}-\mathcal{E}_m^{{\rm M}(h)} 
    + \dfrac{1}{2}\sum\limits_{i,j\in S_n} &\left(J^{\rm M}_{ij,ij} - K^{\rm M}_{ij,ji}\right)
     \text{  if $m = n$}\\
      \dfrac{1}{4} \sum\limits_{j\in S_n} \left(J^{\rm M}_{ij,kj} - K^{\rm M}_{ij,jk}\right) & \text{if $D^M_m$ and $D^M_n$ differ by one single-particle state}\\
      \dfrac{1}{4} \left(J^{\rm M}_{ij,kl} - K^{\rm M}_{ij,lk}\right) & \text{if $D^M_m$ and $D^M_n$ differ by two single-particle states}
    \end{cases}
  \end{split}
\end{equation}
where the factors before the sums in above equation corresponds to $\mathcal{N}$, mentioned in the main paper. Note that the above matrix elements correspond to a diagonal and one of the triangles of the CI matrix. The other triangular matrix is a complex conjugate of the above offdiagonal elements.

\subsection{Method of calculation of configuration interaction}

The sixfold integral in Eq.~\eqref{eq:CoulombMatrElem} is evaluated using the~Green's function method.~\cite{Schliwa:09,Klenovsky2017} The integral in Eq.~\eqref{eq:CoulombMatrElem} is split into solution of the Poisson's equation for one quasiparticle $a$ only, followed by a three-fold integral for quasiparticle $b$ in the electrostatic potential generated by particle $a$ and resulting from the previous step. That procedure, thus, makes the whole solution numerically more feasible and is described by
\begin{equation}
\begin{split}
    \nabla \left[ \epsilon(\mathbf{r}_1) \nabla \hat{U}_{ajl}(\mathbf{r}_1) \right] &= \frac{4\pi e^2}{\epsilon_0}\Psi^*_{aj}(\mathbf{r}_1)\Psi_{al}(\mathbf{r}_1),\\
    V^{{M}}_{ij,kl} &= \int {\rm d}\mathbf{r}_2\,\hat{U}_{ajl}(\mathbf{r}_2)\Psi^*_{bi}(\mathbf{r}_2)\Psi_{bk}(\mathbf{r}_2),
\end{split}
\label{eq:GreenPoisson}
\end{equation}
where $a,b \in \{e,h\}$.

\subsection{Radiative rate \& lifetime}

During computation of CI, the dimensionless oscillator strength of the optical transition $f^{\rm M}_{i}$ of $i$-th state of complex $M$ evaluated using the Fermi's golden rule employing the Hellmann-Feynman theorem~\cite{Dirac1927,Stier1999,Andrzejewski2010,Gaweczyk2017}
\begin{equation}
\label{eq:CIOscStrengthPol}
f^{\rm M}_{i}=\frac{1}{E_i}\frac{2m_0}{\hbar^2}\left|\sum_{\mathit m=1}^{n_{\rm SD}}\eta_{i,m}\sum_{r,p\,\in D_m^{\rm M}}\frac{\left<\psi_{h_r}\left|\hat{\bf e}\cdot\hat{{\bf p}}\right|\psi_{e_p}\right>}{N_{e-h}}\right|^2,
\end{equation}
where $\psi_e$ and $\psi_h$ are the single-particle wavefunctions for electrons and holes, respectively, $\hat{\bf e}$ is the polarization vector and the effect of $\hat{{\bf p}}={\partial\hat{H}_0^{k\cdot p}}/{\partial{\bf k}}$ reduces for the case of interband transitions to overlap between $s$- and $p$-type Bloch components of the corresponding single-particle wavefunctions, multiplied by Kane's parameter $P$~\cite{Stier1999}. Furthermore, $N_{e-h}$ is the normalization factor that reflects the number of summands.

The decay rate of the complex $M$ in homogeneous medium can be obtained from $f^{\rm M}_{i}$ in Eq.~\eqref{eq:CIOscStrengthPol} by~\cite{Stobbe2012}
\begin{equation}
\label{eq:CIemissionRate}
\Gamma^{\rm M}_{i}(E_i)=f^{\rm M}_{i}\Gamma_{cl}(E_i),
\end{equation}
where $\Gamma_{cl}(E_i)$ marks the rate of the classical harmonic oscillator
\begin{equation}
\label{eq:classicalOscillatorOS}    
\Gamma_{cl}(E_i)=\frac{nq^2E_i^2}{6\pi m_0\epsilon_0\hbar^2c^3},
\end{equation}
where $n$ is the refractive index of the medium, $q$ is the elementary charge, $m_0$ the free electron mass, $\epsilon_0$ the vacuum dielectric constant, $\hbar$ the reduced Planck's constant and $c$ is the speed of light.
Note that the values of radiative lifetime discussed in the main paper are obtained as an inverse of $\Gamma^{\rm M}_{i}(\omega)$ in Eq.~\eqref{eq:CIemissionRate}.

\section{Additional theory results}    
	The structure of the QD is depicted in Fig.~\ref{Fig:Structure}~(d). In our theoretical analysis, we vary the QD height (H) and base length (W) to capture the effect arising from QD shape changes on the simulated properties. Additionally, the indium content ($x$) within the QDs is adjusted to explore its influence on the system's behavior and its impact on the emission characteristics of the emitter. Throughout all calculations, the WL thickness and composition remains constant, ensuring that any observed effects stem solely from modifications to the QD height, base length, and composition. 
    
		\begin{figure*}[t]  
		\centering
        \renewcommand{\thefigure}{S6}
		\includegraphics[width=\textwidth]{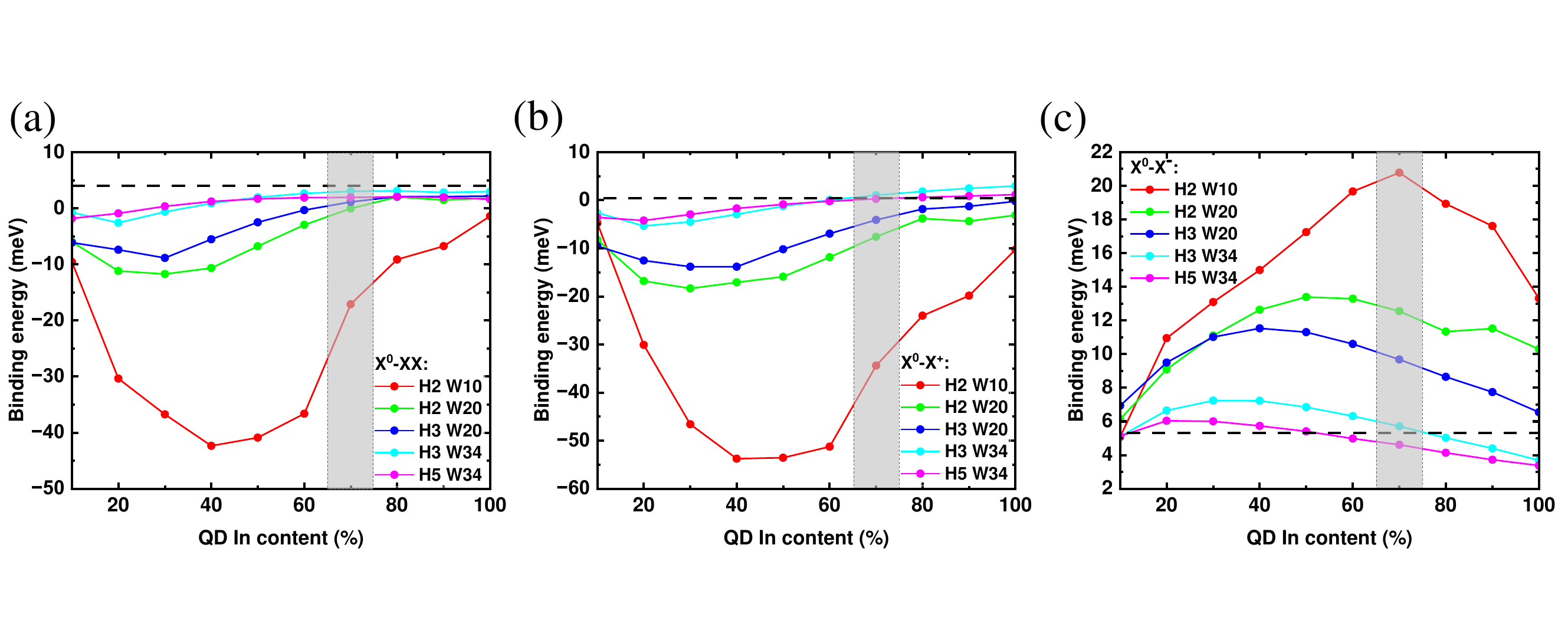}  
		\caption{The simulated binding energies of the biexciton ($XX$), positively charged trion ($X^+$), and negatively charged trion (X$^-$), shown in panels (a), (b), and (c), respectively, are given relative to the exciton ($X^0$) energy, which is set to 0~meV, for various QD sizes. The gray box indicates the range where the experimental data and theoretical model show the best agreement, corresponding to an indium concentration of \(70 \pm 5\%\). The multiparticle calculations shown here were performed with single-particle basis of twelve electron and twenty-four hole states and using the singles-doubles approximation to enable CI calculation with such a large basis numerically feasible.}
		\label{Fig:MultipartBinding}
	\end{figure*}
    In Fig.~\ref{Fig:MultipartBinding} we show the evolution of the binding energies of biexciton ($XX$), positive ($X^+$) and negative (X$^-$) trions with increasing indium content ($x$) for several sizes of the studied In$_x$Ga$_{1-x}$As QD. The complexes were computed using singles-doubles CI (SDCI)~\cite{Schliwa:09} with the basis of twelve single-particle electron and twenty-four single-particle hole states. The large basis size ensured that most of the effect of Coulomb correlation was included in the calculations. On the other hand, the SDCI approximation ensured numerical feasibility of the calculations.

    We see in Fig.~\ref{Fig:MultipartBinding}~(a)~and~(b) that $XX$ and $X^+$ are anti-binding for the smallest computed dots. However, these complexes become binding for increased dot size and indium content, reaching close to experimental observations for QD with height of 3~nm, basis length of 34~nm and indium content of 70\%. In Fig.~\ref{Fig:MultipartBinding}~(c) we see that $X^-$ is found binding for all studied dot sizes and compositions, in agreement with previous studies~\cite{Schliwa:09}.Similarly, as for $XX$ and $X^+$, the best agreement of the binding energy of $X^-$ with our experiments is reached for 3~nm high QD with basis length of 34~nm and 70\% indium content.

\begin{figure*}[t]  
	\centering
    \renewcommand{\thefigure}{S7}
	\includegraphics[width=\textwidth]{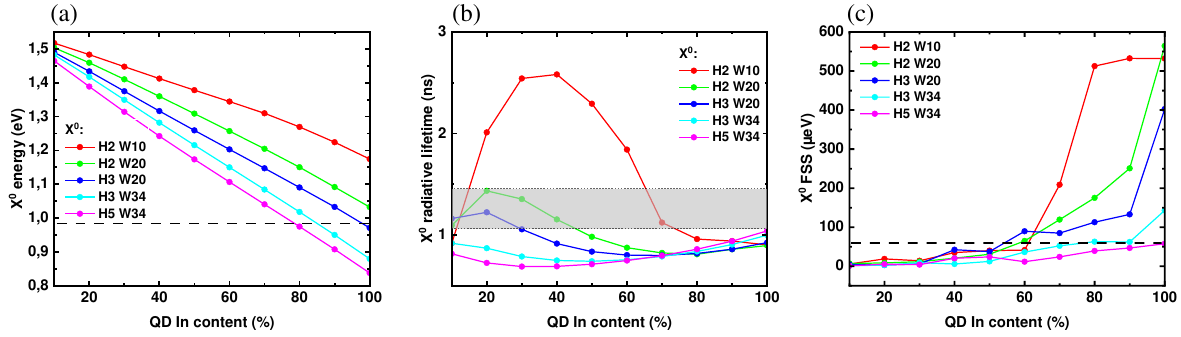}  
	\caption{(a)~The simulated exciton (X) energy is shown for QDs of different sizes as a function of the indium content. The black dotted line indicates the experimentally measured energy of the $X$ transition at 4~K. (b)~The radiative lifetime of the exciton (X) transition is calculated for various QD sizes as a function of the indium concentration. The gray box represents the experimentally measured value with its associated fitting uncertainty. The CI calculations of excitons were performed with the basis size of twenty-four electron and hole single-particle states,~i.e. the effect of the Coulomb correlation was included almost completely~\cite{Schliwa:09}. (c) Simulated FSS of the X$^0$ line as a function of varying indium content and different QD sizes. The calculation of FSS involves the multipole expansion of the exchange interaction~\cite{Takagahara2000,Krapek2015}. Calculations in were performed by CI with symmetric basis, i.e., with the same number of twenty-four electron and twenty-four hole single-particle states.}
	\label{Fig:ExcitonEL}
    \end{figure*}

    The evolution of the computed exciton FSS with increasing indium content in QD is shown in Fig.~\ref{Fig:ExcitonEL}~(c). The calculations are given for the same sizes of QDs as in Fig.~\ref{Fig:MultipartBinding}. The FSS calculations involved the effect of the multipole expansion of the exchange interaction~\cite{Takagahara2000,Krapek2015} which is crucial for obtaining realistic FSS values for QDs with larger indium content. That effect stands behind the considerable increase of FSS for QD indium content from ca 70\% to 100\% as seen in Fig.~\ref{Fig:ExcitonEL}~(c) for all studied QD shapes.

    The evolution of the exciton energies and lifetimes with indium content for different QD sizes is shown in Fig.~\ref{Fig:ExcitonEL}~(a)~and~(b), respectively. We see that best agreement with experiment for exciton energy is reached for the dots with 3~nm height and 34~nm base length and that with height of 5~nm and same base length. Satisfactorily, the agreement with experiment for both aforementioned dots is reached for indium concentration of 70\%-80\% similarly as for binding energies of the multi-particle complexes above. However, the agreement with experiment is slightly worse for exciton lifetime, but the difference to that is only on the order of 100-200~ps for the dots with 3~nm height and base length of 34~nm and that with the same base length but 5~nm height. For these dots again the best agreement is reached for 70\% indium content in QD.

    \begin{figure*}[t]  
	\centering
    \renewcommand{\thefigure}{S8}
	\includegraphics[width=\textwidth]{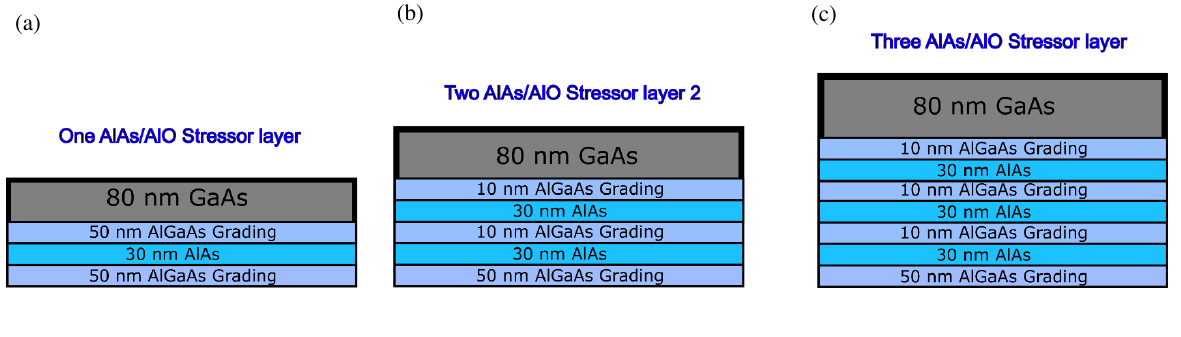}  
	\caption{Layer designs of simulated structures for the multi-buried-stressor approach: (a) structure employed experimentally in this work with a single AlAs/AlO stressor layer, (b) structure with two AlAs/AlO stressor layers, and (c) structure with three AlAs/AlO stressor layers.}
	\label{Fig:Strain_Structure}
    \end{figure*}

In Fig.~\ref{Fig:Strain_Structure} we show the layer design of the multi-stressor structures used for calculations of surface strain in Fig.~4 of the main paper. The calculations of the elastic strain were performed using the continuum elasticity method, discussed elsewhere~\cite{Gaur2025-gw} implemented within Nextnano++~\cite{Birner2007}.

    
\newpage

\clearpage
\section*{References}
\bibliographystyle{unsrt}
\bibliography{References}

@article{Gisin2007,
   author = {Nicolas Gisin and Rob Thew},
   doi = {10.1038/nphoton.2007.22},
   issn = {1749-4885},
   issue = {3},
   journal = {Nature Photonics},
   month = {3},
   pages = {165-171},
   title = {Quantum communication},
   volume = {1},
   year = {2007},
}

@article{Long2007,
   author = {Gui-lu Long and Fu-guo Deng and Chuan Wang and Xi-han Li and Kai Wen and Wan-ying Wang},
   doi = {10.1007/s11467-007-0050-3},
   issn = {1673-3487},
   issue = {3},
   journal = {Frontiers of Physics in China},
   month = {7},
   pages = {251-272},
   title = {Quantum secure direct communication and deterministic secure quantum communication},
   volume = {2},
   year = {2007},
}

@article{Heindel2023,
   author = {Tobias Heindel and Je-Hyung Kim and Niels Gregersen and Armando Rastelli and Stephan Reitzenstein},
   doi = {10.1364/AOP.490091},
   issn = {1943-8206},
   issue = {3},
   journal = {Advances in Optics and Photonics},
   month = {9},
   pages = {613},
   title = {Quantum dots for photonic quantum information technology},
   volume = {15},
   year = {2023},
}

@article{Somaschi2016,
   author = {N. Somaschi and V. Giesz and L. De Santis and J. C. Loredo and M. P. Almeida and G. Hornecker and S. L. Portalupi and T. Grange and C. Antón and J. Demory and C. Gómez and I. Sagnes and N. D. Lanzillotti-Kimura and A. Lemaítre and A. Auffeves and A. G. White and L. Lanco and P. Senellart},
   doi = {10.1038/nphoton.2016.23},
   issn = {1749-4885},
   issue = {5},
   journal = {Nature Photonics},
   month = {5},
   pages = {340-345},
   title = {Near-optimal single-photon sources in the solid state},
   volume = {10},
   year = {2016},
}

@article{Holewa2020,
   author = {Paweł Holewa and Marek Burakowski and Anna Musiał and Nicole Srocka and David Quandt and André Strittmatter and Sven Rodt and Stephan Reitzenstein and Grzegorz Sek},
   doi = {10.1038/s41598-020-78462-4},
   issn = {2045-2322},
   issue = {1},
   journal = {Scientific Reports},
   month = {12},
   pages = {21816},
   title = {Thermal stability of emission from single {InGaAs/GaAs} quantum dots at the telecom {O}-band},
   volume = {10},
   year = {2020},
}

@article{Strittmatter2012,
   author = {André Strittmatter and André Holzbecher and Andrei Schliwa and Jan-Hindrik Schulze and David Quandt and Tim David Germann and Alexander Dreismann and Ole Hitzemann and Erik Stock and Irina A. Ostapenko and Sven Rodt and Waldemar Unrau and Udo W. Pohl and Axel Hoffmann and Dieter Bimberg and Vladimir Haisler},
   doi = {10.1002/pssa.201228407},
   issn = {18626300},
   issue = {12},
   journal = {physica status solidi (a)},
   month = {12},
   pages = {2411-2420},
   title = {Site-controlled quantum dot growth on buried oxide stressor layers},
   volume = {209},
   year = {2012},
}

@article{Limame2024b,
   author = {Imad Limame and Ching-Wen Shih and Alexej Koltchanov and Fabian Heisinger and Felix Nippert and Moritz Plattner and Johannes Schall and Markus R. Wagner and Sven Rodt and Petr Klenovsky and Stephan Reitzenstein},
   doi = {10.1063/5.0187074},
   issn = {0003-6951},
   issue = {6},
   journal = {Applied Physics Letters},
   month = {2},
   title = {Epitaxial growth and characterization of multi-layer site-controlled {InGaAs} quantum dots based on the buried stressor method},
   volume = {124},
   year = {2024},
}

@book{Ustinov2003,
   author = {Victor M. Ustinov and Alexey E. Zhukov and Anton Yu. Egorov and Nikolai A. Maleev},
   doi = {10.1093/acprof:oso/9780198526797.001.0001},
   isbn = {0198526792},
   month = {8},
   publisher = {Oxford University PressOxford},
   title = {Quantum Dot Lasers},
   year = {2003},
}

@article{Kaganskiy2019,
   author = {Arsenty Kaganskiy and Sören Kreinberg and Xavier Porte and Stephan Reitzenstein},
   doi = {10.1364/OPTICA.6.000404},
   issn = {2334-2536},
   issue = {4},
   journal = {Optica},
   month = {4},
   pages = {404},
   title = {Micropillar lasers with site-controlled quantum dots as active medium},
   volume = {6},
   year = {2019},
}

@article{Shih2024,
   author = {Ching‐Wen Shih and Imad Limame and Chirag C. Palekar and Aris Koulas‐Simos and Arsenty Kaganskiy and Petr Klenovský and Stephan Reitzenstein},
   doi = {10.1002/lpor.202301242},
   issn = {1863-8880},
   journal = {Laser \& Photonics Reviews},
   month = {3},
   title = {Self‐Aligned Photonic Defect Microcavity Lasers with Site‐Controlled Quantum Dots},
   year = {2024},
}

@article{Nawrath2019,
   author = {C. Nawrath and F. Olbrich and M. Paul and S. L. Portalupi and M. Jetter and P. Michler},
   doi = {10.1063/1.5095196},
   issn = {0003-6951},
   issue = {2},
   journal = {Applied Physics Letters},
   month = {7},
   title = {Coherence and indistinguishability of highly pure single photons from non-resonantly and resonantly excited telecom {C}-band quantum dots},
   volume = {115},
   year = {2019},
}

@article{Musial2020,
   author = {Anna Musia\l{} and Kinga \.{Z}o\l{}nacz and Nicole Srocka and Oleh Kravets and Jan Gro\ss{}e and Jacek Olszewski and Krzysztof Poturaj and Grzegorz W\'{o}jcik and Pawe\l{} Mergo and Kamil Dybka and Mariusz Dyrkacz and Micha\l{} D\l{}ubek and Kristian Lauritsen and Andreas B\"{u}lter and Philipp--Immanuel Schneider and Lin Zschiedrich and Sven Burger and Sven Rodt and Wac\l{}aw Urbanczyk and Grzegorz Sek and Stephan Reitzenstein},
   doi = {10.1002/qute.202000018},
   issn = {2511-9044},
   issue = {6},
   journal = {Advanced Quantum Technologies},
   month = {6},
   title = {Plug \& Play Fiber‐Coupled 73 {kHz} Single‐Photon Source Operating in the Telecom {O}‐Band},
   volume = {3},
   year = {2020},
}

@article{Gao2022,
   author = {Timm Gao and Lucas Rickert and Felix Urban and Jan Gro\ss{}e and Nicole Srocka and Sven Rodt and Anna Musia\l{} and Kinga \.{Z}o\l{}nacz and Pawe\l{} Mergo and Kamil Dybka and Wac\l{}aw Urbanczyk and Grzegorz Sek and Sven Burger and Stephan Reitzenstein and Tobias Heindel},
   doi = {10.1063/5.0070966},
   issn = {1931-9401},
   issue = {1},
   journal = {Applied Physics Reviews},
   month = {3},
   title = {A quantum key distribution testbed using a plug \& play telecom-wavelength single-photon source},
   volume = {9},
   year = {2022},
}

@article{Schlehahn2018,
   author = {Alexander Schlehahn and Sarah Fischbach and Ronny Schmidt and Arsenty Kaganskiy and André Strittmatter and Sven Rodt and Tobias Heindel and Stephan Reitzenstein},
   doi = {10.1038/s41598-017-19049-4},
   issn = {2045-2322},
   issue = {1},
   journal = {Scientific Reports},
   month = {1},
   pages = {1340},
   title = {A stand-alone fiber-coupled single-photon source},
   volume = {8},
   year = {2018},
}

@article{Denning2020,
   author = {Emil V. Denning and Jake Iles-Smith and Niels Gregersen and Jesper Mork},
   doi = {10.1364/OME.380601},
   issn = {2159-3930},
   issue = {1},
   journal = {Optical Materials Express},
   month = {1},
   pages = {222},
   title = {Phonon effects in quantum dot single-photon sources},
   volume = {10},
   year = {2020},
}

@article{Braun2014,
   author = {T. Braun and C. Schneider and S. Maier and R. Igusa and S. Iwamoto and A. Forchel and S. Höfling and Y. Arakawa and M. Kamp},
   doi = {10.1063/1.4896284},
   issn = {2158-3226},
   issue = {9},
   journal = {AIP Advances},
   month = {9},
   title = {Temperature dependency of the emission properties from positioned {In(Ga)As/GaAs} quantum dots},
   volume = {4},
   year = {2014},
}

@article{Xu2008,
   author = {Zhangcheng Xu and Yating Zhang and Jørn M. Hvam},
   doi = {10.1063/1.3021018},
   issn = {0003-6951},
   issue = {18},
   journal = {Applied Physics Letters},
   month = {11},
   title = {Long luminescence lifetime in self-assembled {InGaAs/GaAs} quantum dots at room temperature},
   volume = {93},
   year = {2008},
}

@article{Olbrich2017,
   author = {Fabian Olbrich and Jan Kettler and Matthias Bayerbach and Matthias Paul and Jonatan Höschele and Simone Luca Portalupi and Michael Jetter and Peter Michler},
   doi = {10.1063/1.4983362},
   issn = {0021-8979},
   issue = {18},
   journal = {Journal of Applied Physics},
   month = {5},
   title = {Temperature-dependent properties of single long-wavelength {InGaAs} quantum dots embedded in a strain reducing layer},
   volume = {121},
   year = {2017},
}

@article{Dusanowski2014,
   author = {\L{} Dusanowski and A. Musial and A. Marynski and P. Mrowinski and J. Andrzejewski and P. Machnikowski and J. Misiewicz and A. Somers and S. H\"{o}fling and J. P. Reithmaier and G. Sek },
   doi = {10.1103/PhysRevB.90.125424},
   issn = {1098-0121},
   issue = {12},
   journal = {Physical Review B},
   month = {9},
   pages = {125424},
   title = {Phonon-assisted radiative recombination of excitons confined in strongly anisotropic nanostructures},
   volume = {90},
   year = {2014},
}

@article{Ortner2005,
   author = {G. Ortner and M. Schwab and M. Bayer and R. Pässler and S. Fafard and Z. Wasilewski and P. Hawrylak and A. Forchel},
   doi = {10.1103/PhysRevB.72.085328},
   issn = {1098-0121},
   issue = {8},
   journal = {Physical Review B},
   month = {8},
   pages = {085328},
   title = {Temperature dependence of the excitonic band gap in $\mathrm{In}_x\mathrm{Ga}_{1-x}\mathrm{As}/\mathrm{GaAs}$   self-assembled quantum dots},
   volume = {72},
   year = {2005},
}

@Article{Krapek2015,
  author  = {K\v{r}\'apek, V. and Klenovsk\'y, P. and \v{S}ikola, T.},
  title   = {Excitonic fine structure splitting in type-{II} quantum dots},
  journal = {Physical Review B},
  year    = {2015},
  volume  = {92},
  number  = {19},
  pages   = {195430},
  month   = nov,
  ddoi     = {10.1103/PhysRevB.92.195430},
  ei      = {1550-235X},
  ri      = {Krapek, Vlastimil/A-6917-2013; Sikola, Tomas/D-9875-2012},
  sn      = {1098-0121},
  tc      = {7},
  ut      = {WOS:000365507900004},
  z8      = {0},
  z9      = {7},
  zb      = {1},
  zr      = {0},
  zs      = {0},
}

@article{Takagahara2000,
author = {Takagahara, T.},
ddoi = {10.1103/PhysRevB.62.16840},
issn = {01631829},
journal = {Physical Review B},
month = {dec},
number = {24},
pages = {16840},
publisher = {American Physical Society},
title = {{Theory of exciton doublet structures and polarization relaxation in single quantum dots}},
url = {https://journals.aps.org/prb/abstract/10.1103/PhysRevB.62.16840},
volume = {62},
year = {2000}
}

@Article{Bester:06,
  Title                    = {Importance of Second-Order Piezoelectric Effects in Zinc-Blende Semiconductors},
  Author                   = {Bester, Gabriel and Wu, Xifan and Vanderbilt, David and Zunger, Alex},
  Journal                  = {Physical Review Letters},
  Year                     = {2006},

  Month                    = {May},
  Pages                    = {187602},
  Volume                   = {96},

  ddoi                      = {10.1103/PhysRevLett.96.187602},
  Issue                    = {18},
  Numpages                 = {4},
  Publisher                = {American Physical Society},
  Url                      = {http://link.aps.org/ddoi/10.1103/PhysRevLett.96.187602}
}

@Article{Beya-Wakata2011,
  Title                    = {First- and second-order piezoelectricity in {III-V} semiconductors},
  Author                   = {Beya-Wakata, A. and Prodhomme, P. Y. and Bester, G.},
  Journal                  = {Physical Review B},
  Year                     = {2011},

  Month                    = nov,
  Number                   = {19},
  Pages                    = {195207},
  Volume                   = {84},

  Af                       = {Beya-Wakata, AnnieEOLEOLProdhomme, Pierre-YvesEOLEOLBester, Gabriel},
  ddoi                      = {10.1103/PhysRevB.84.195207},
  Ei                       = {2469-9969},
  Sn                       = {2469-9950},
  Tc                       = {31},
  Ut                       = {WOS:000297414500027},
  Z9                       = {31}
}

@article{Klenovsky2018,
author = {Klenovsk{\'{y}}, Petr and Steindl, Petr and Aberl, Johannes and Zallo, Eugenio and Trotta, Rinaldo and Rastelli, Armando and Fromherz, Thomas},
ddoi = {10.1103/PhysRevB.97.245314},
issn = {24699969},
journal = {Physical Review B},
number = {24},
pages = {245314},
title = {{Effect of second-order piezoelectricity on the excitonic structure of stress-tuned {In(Ga)As/GaAs} quantum dots}},
volume = {97},
year = {2018}
}

@article{Birner2007,
author = {Birner, Stefan and Zibold, Tobias and Andlauer, Till and Kubis, Tillmann and Sabathil, Matthias and Trellakis, Alex and Vogl, Peter},
ddoi = {10.1109/TED.2007.902871},
issn = {00189383},
journal = {IEEE Transactions on Electron Devices},
month = {sep},
number = {9},
pages = {2137--2142},
title = {{Nextnano: General purpose 3-{D} simulations}},
volume = {54},
year = {2007}
}

@article{Klenovsky2017,
  title = {Excitonic structure and pumping power dependent emission blue-shift of type-{II} quantum dots},
  author = {Klenovsk\'y, P. and Steindl, P. and Geffroy, D.},
  journal = {Scientific Reports},
  volume = {7},
  issue = {1},
  pages = {45568},
  numpages = {0},
  year = {2017},
  month = {Mar},
  publisher = {American Physical Society},
}

@Article{Schliwa:09,
  author    = {Schliwa, Andrei and Winkelnkemper, Momme and Bimberg, Dieter},
  title = {Few-Particle Energies versus Geometry and Composition of {In$_x$Ga$_{1-x}$As/GaAs} Self-Organized Quantum Dots},
  journal   = {Physical Review B},
  year      = {2009},
  volume    = {79},
  pages     = {075443},
  month     = {Feb},
  ddoi       = {10.1103/PhysRevB.79.075443},
  issue     = {7},
  numpages  = {14},
  publisher = {American Physical Society},
  url       = {http://link.aps.org/ddoi/10.1103/PhysRevB.79.075443},
}

@article{Dirac1927,
author = {Dirac, PAM},
ddoi = {10.1098/RSPA.1927.0039},
issn = {0950-1207},
journal = {Proceedings of the Royal Society of London. Series A, Containing Papers of a Mathematical and Physical Character},
month = {mar},
number = {767},
pages = {243--265},
publisher = {The Royal Society London},
title = {{The quantum theory of the emission and absorption of radiation}},
url = {https://royalsocietypublishing.org/ddoi/abs/10.1098/rspa.1927.0039},
volume = {114},
year = {1927}
}

@article{Andrzejewski2010,
   author = {Janusz Andrzejewski and Grzegorz Sek. and Eoin O'Reilly and Andrea Fiore and Jan Misiewicz},
   doi = {10.1063/1.3346552/380135},
   issn = {00218979},
   issue = {7},
   journal = {Journal of Applied Physics},
   month = {4},
   publisher = {AIP Publishing},
   title = {Eight-band k.p calculations of the composition contrast effect on the linear polarization properties of columnar quantum dots},
   volume = {107},
   url = {/aip/jap/article/107/7/073509/380135/Eight-band-k-p-calculations-of-the-composition},
   year = {2010}
}

@article{Gaweczyk2017,
   author = {M. Gawełczyk and M. Syperek and A. Maryński and P. Mrowiński and Dusanowski and K. Gawarecki and J. Misiewicz and A. Somers and J. P. Reithmaier and S. Höfling and G. Sek},
   doi = {10.1103/PHYSREVB.96.245425/FIGURES/8/THUMBNAIL},
   issn = {24699969},
   issue = {24},
   journal = {Physical Review B},
   month = {12},
   pages = {245425},
   publisher = {American Physical Society},
   title = {Exciton lifetime and emission polarization dispersion in strongly in-plane asymmetric nanostructures},
   volume = {96},
   url = {https://journals.aps.org/prb/abstract/10.1103/PhysRevB.96.245425},
   year = {2017}
}

@article{Stier1999,
   author = {O. Stier and M. Grundmann and D. Bimberg},
   doi = {10.1103/PhysRevB.59.5688},
   issn = {1550235X},
   issue = {8},
   journal = {Physical Review B},
   month = {2},
   pages = {5688},
   publisher = {American Physical Society},
   title = {Electronic and optical properties of strained quantum dots modeled by 8-band k.p theory},
   volume = {59},
   url = {https://journals.aps.org/prb/abstract/10.1103/PhysRevB.59.5688},
   year = {1999}
}

@article{Stobbe2012,
   author = {S. Stobbe and P. T. Kristensen and J. E. Mortensen and J. M. Hvam and J. Mørk and P. Lodahl},
   dddoi = {10.1103/PhysRevB.86.085304},
   issn = {1098-0121},
   issue = {8},
   journal = {Physical Review B},
   month = {8},
   pages = {085304},
   title = {Spontaneous emission from large quantum dots in nanostructures: Exciton-photon interaction beyond the dipole approximation},
   volume = {86},
   year = {2012},
}

@article{Bahder1990,
   author = {Thomas B. Bahder},
   doi = {10.1103/PhysRevB.41.11992},
   issn = {01631829},
   issue = {17},
   journal = {Physical Review B},
   month = {6},
   pages = {11992},
   publisher = {American Physical Society},
   title = {Eight-band k.p model of strained zinc-blende crystals},
   volume = {41},
   url = {https://journals.aps.org/prb/abstract/10.1103/PhysRevB.41.11992},
   year = {1990}
}

@article{Bryant1987,
   author = {Garnett W. Bryant},
   doi = {10.1103/PhysRevLett.59.1140},
   issn = {00319007},
   issue = {10},
   journal = {Physical Review Letters},
   month = {9},
   pages = {1140},
   publisher = {American Physical Society},
   title = {Electronic structure of ultrasmall quantum-well boxes},
   volume = {59},
   url = {https://journals.aps.org/prl/abstract/10.1103/PhysRevLett.59.1140},
   year = {1987}
}

@article{Troparevsky2008,
   author = {M. C. Troparevsky and A. Franceschetti},
   dddoi = {10.1088/0953-8984/20/5/055211},
   issn = {0953-8984},
   issue = {5},
   journal = {Journal of Physics: Condensed Matter},
   month = {1},
   pages = {055211},
   publisher = {IOP Publishing},
   title = {An optimized configuration interaction method for calculating electronic excitations innanostructures},
   volume = {20},
   url = {https://iopscience.iop.org/article/10.1088/0953-8984/20/5/055211 https://iopscience.iop.org/article/10.1088/0953-8984/20/5/055211/meta},
   year = {2008},
}

@ARTICLE{Ries2024-cx,
  title     = "Raman-imaging of the strain-distribution in semiconductors",
  author    = "Ries, Maximilian and Heisinger, Fabian and Limame, Imad",
  journal   = "BIO Web Conf.",
  publisher = "EDP Sciences",
  volume    =  129,
  pages     = "06031",
  year      =  2024,
  copyright = "https://creativecommons.org/licenses/by/4.0/"
}

@ARTICLE{Ries2024-gp,
  title     = "Molecular microscopy by Thermo-fisher-scientific: {FTIR-imaging}
               of a Tintoretto-fresco and Raman-imaging of the
               strain-distribution in semiconductors",
  author    = "Ries, Maximilian and Bravo, Barbara and Balliana, Eleonora and
               Heisinger, Fabian and Limame, Imad",
  journal   = "BIO Web Conf.",
  publisher = "EDP Sciences",
  volume    =  129,
  pages     = "06022",
  year      =  2024,
  copyright = "https://creativecommons.org/licenses/by/4.0/"
}

@misc{Millington-Hotze2025,
  doi = {10.48550/ARXIV.2504.19257},
  url = {https://arxiv.org/abs/2504.19257},
  author = {Millington-Hotze,  Peter and Klenovsky,  Petr and Dyte,  Harry E. and Gillard,  George and Manna,  Santanu and da Silva,  Saimon F. Covre and Rastelli,  Armando and Chekhovich,  Evgeny A.},
  title = {Few-electron spin qubits in optically active GaAs quantum dots},
  publisher = {arXiv},
  year = {2025},
  copyright = {Creative Commons Attribution 4.0 International}
}

@article{Hauser2025-hx,
  title = {Deterministic and highly indistinguishable single photons in the telecom {C}-band},
  volume = {17},
  ISSN = {2041-1723},
  url = {http://dx.doi.org/10.1038/s41467-026-68336-0},
  DOI = {10.1038/s41467-026-68336-0},
  number = {1},
  journal = {Nature Communications},
  publisher = {Springer Science and Business Media LLC},
  author = {Hauser,  Nico and Bayerbach,  Matthias and Kaupp,  Jochen and Reum,  Yorick and Peniakov,  Giora and Michl,  Johannes and Kamp,  Martin and Huber-Loyola,  Tobias and Pfenning,  Andreas T. and H\"{o}fling,  Sven and Barz,  Stefanie},
  year = {2026},
  month = jan 
}

@ARTICLE{Grose2020-qa,
  title     = "Development of site-controlled quantum dot arrays acting as
               scalable sources of indistinguishable photons",
  author    = "Gro{\ss}e, Jan and von Helversen, Martin and Koulas-Simos, Aris
               and Hermann, Martin and Reitzenstein, Stephan",
  journal   = "APL Photonics",
  publisher = "AIP Publishing",
  volume    =  5,
  number    =  9,
  pages     = "096107",
  month     =  sep,
  year      =  2020,
  copyright = "https://creativecommons.org/licenses/by/4.0/",
  language  = "en"
}

@ARTICLE{Hanschke2018,
  title     = "Quantum dot single-photon sources with ultra-low multi-photon
               probability",
  author    = "Hanschke, Lukas and Fischer, Kevin A and Appel, Stefan and
               Lukin, Daniil and Wierzbowski, Jakob and Sun, Shuo and Trivedi,
               Rahul and Vu{\v c}kovi{\'c}, Jelena and Finley, Jonathan J and
               M{\"u}ller, Kai",
  journal   = "Npj Quantum Inf.",
  publisher = "Springer Science and Business Media LLC",
  volume    =  4,
  number    =  1,
  month     =  sep,
  year      =  2018,
  copyright = "https://creativecommons.org/licenses/by/4.0",
  language  = "en"
}

@ARTICLE{Margaria2025-tz,
  title     = "Efficient fibre-pigtailed source of indistinguishable single
               photons",
  author    = "Margaria, Nico and Pastier, Florian and Bennour, Thinhinane and
               Billard, Marie and Ivanov, Edouard and Hease, William and
               Stepanov, Petr and Adiyatullin, Albert F and Singla, Raksha and
               Pont, Mathias and Descampeaux, Maxime and Bernard, Alice and
               Pishchagin, Anton and Morassi, Martina and Lema{\^\i}tre,
               Aristide and Volz, Thomas and Giesz, Val{\'e}rian and Somaschi,
               Niccolo and Maring, Nicolas and Boissier, S{\'e}bastien and Au,
               Thi Huong and Senellart, Pascale",
  journal   = "Nat. Commun.",
  publisher = "Springer Science and Business Media LLC",
  volume    =  16,
  number    =  1,
  pages     = "7553",
  month     =  aug,
  year      =  2025,
  copyright = "https://creativecommons.org/licenses/by/4.0",
  language  = "en"
}

@ARTICLE{Paul2017-ef,
  title     = "Single-photon emission at 1.55 \textit{$\mu$}m from
               {MOVPE-grown} {InAs} quantum dots on {InGaAs/GaAs} metamorphic
               buffers",
  author    = "Paul, Matthias and Olbrich, Fabian and H{\"o}schele, Jonatan and
               Schreier, Susanne and Kettler, Jan and Portalupi, Simone Luca
               and Jetter, Michael and Michler, Peter",
  journal   = "Appl. Phys. Lett.",
  publisher = "AIP Publishing",
  volume    =  111,
  number    =  3,
  pages     = "033102",
  month     =  jul,
  year      =  2017,
  language  = "en"
}

@ARTICLE{Paul2015-jz,
  title     = "Metal-organic vapor-phase epitaxy-grown ultra-low density
               {InGaAs/GaAs} quantum dots exhibiting cascaded single-photon
               emission at 1.3 \textit{$\mu$}m",
  author    = "Paul, Matthias and Kettler, Jan and Zeuner, Katharina and
               Clausen, Caterina and Jetter, Michael and Michler, Peter",
  journal   = "Appl. Phys. Lett.",
  publisher = "AIP Publishing",
  volume    =  106,
  number    =  12,
  pages     = "122105",
  month     =  mar,
  year      =  2015,
  language  = "en"
}

@ARTICLE{Gaur2025-gw,
  title     = "Buried-stressor technology for the epitaxial growth and device
               integration of site-controlled quantum dots",
  author    = "Gaur, Kartik and Mudi, Priyabrata and Klenovsky, Petr and
               Reitzenstein, Stephan",
  journal   = "Mater. Quantum Technol.",
  publisher = "IOP Publishing",
  volume    =  5,
  number    =  2,
  pages     = "022002",
  month     =  jun,
  year      =  2025,
  copyright = "https://creativecommons.org/licenses/by/4.0/"
}

@ARTICLE{Stock2011-ad,
  title     = "Acoustic and optical phonon scattering in a single {In(Ga)As}
               quantum dot",
  author    = "Stock, Erik and Dachner, Matthias-Rene and Warming, Till and
               Schliwa, Andrei and Lochmann, Anatol and Hoffmann, Axel and
               Toropov, Aleksandr I and Bakarov, Askhat K and Derebezov, Ilya A
               and Richter, Marten and Haisler, Vladimir A and Knorr, Andreas
               and Bimberg, Dieter",
  journal   = "Phys. Rev. B Condens. Matter Mater. Phys.",
  publisher = "American Physical Society (APS)",
  volume    =  83,
  number    =  4,
  month     =  jan,
  year      =  2011,
  copyright = "http://link.aps.org/licenses/aps-default-license"
}

@ARTICLE{Wang2019-vw,
  title     = "On-demand semiconductor source of entangled photons which
               simultaneously has high fidelity, efficiency, and
               indistinguishability",
  author    = "Wang, Hui and Hu, Hai and Chung, T-H and Qin, Jian and Yang,
               Xiaoxia and Li, J-P and Liu, R-Z and Zhong, H-S and He, Y-M and
               Ding, Xing and Deng, Y-H and Dai, Qing and Huo, Y-H and
               H{\"o}fling, Sven and Lu, Chao-Yang and Pan, Jian-Wei",
  journal   = "Phys. Rev. Lett.",
  publisher = "American Physical Society (APS)",
  volume    =  122,
  number    =  11,
  pages     = "113602",
  month     =  mar,
  year      =  2019,
  copyright = "https://link.aps.org/licenses/aps-default-license",
  language  = "en"
}

@ARTICLE{Lettner2021-ce,
  title     = "Strain-controlled quantum dot fine structure for entangled
               photon generation at 1550 nm",
  author    = "Lettner, Thomas and Gyger, Samuel and Zeuner, Katharina D and
               Schweickert, Lucas and Steinhauer, Stephan and Reuterski{\"o}ld
               Hedlund, Carl and Stroj, Sandra and Rastelli, Armando and
               Hammar, Mattias and Trotta, Rinaldo and J{\"o}ns, Klaus D and
               Zwiller, Val",
  journal   = "Nano Lett.",
  publisher = "American Chemical Society (ACS)",
  volume    =  21,
  number    =  24,
  pages     = "10501--10506",
  month     =  dec,
  year      =  2021,
  copyright = "https://creativecommons.org/licenses/by/4.0/",
  language  = "en"
}

@ARTICLE{Kolatschek2021-zz,
  title     = "Bright Purcell enhanced single-photon source in the telecom
               {O}-band based on a quantum dot in a circular Bragg grating",
  author    = "Kolatschek, Sascha and Nawrath, Cornelius and Bauer, Stephanie
               and Huang, Jiasheng and Fischer, Julius and Sittig, Robert and
               Jetter, Michael and Portalupi, Simone Luca and Michler, Peter",
  journal   = "Nano Lett.",
  publisher = "American Chemical Society (ACS)",
  volume    =  21,
  number    =  18,
  pages     = "7740--7745",
  month     =  sep,
  year      =  2021,
  copyright = "https://creativecommons.org/licenses/by/4.0/",
  language  = "en"
}

@ARTICLE{Schneider2009-lb,
  title     = "Single photon emission from a site-controlled quantum
               dot-micropillar cavity system",
  author    = "Schneider, C and Heindel, T and Huggenberger, A and Weinmann, P
               and Kistner, C and Kamp, M and Reitzenstein, S and H{\"o}fling,
               S and Forchel, A",
  journal   = "Appl. Phys. Lett.",
  publisher = "AIP Publishing",
  volume    =  94,
  number    =  11,
  pages     = "111111",
  month     =  mar,
  year      =  2009,
  language  = "en"
}

@ARTICLE{Jons2013-ky,
  title     = "Triggered indistinguishable single photons with narrow line
               widths from site-controlled quantum dots",
  author    = "J{\"o}ns, K D and Atkinson, P and M{\"u}ller, M and Heldmaier, M
               and Ulrich, S M and Schmidt, O G and Michler, P",
  journal   = "Nano Lett.",
  publisher = "American Chemical Society (ACS)",
  volume    =  13,
  number    =  1,
  pages     = "126--130",
  month     =  jan,
  year      =  2013,
  language  = "en"
}

@ARTICLE{Haffouz2018-ql,
  title     = "Bright single {InAsP} quantum dots at telecom wavelengths in
               position-controlled {InP} nanowires: The role of the photonic
               waveguide",
  author    = "Haffouz, Sofiane and Zeuner, Katharina D and Dalacu, Dan and
               Poole, Philip J and Lapointe, Jean and Poitras, Daniel and
               Mnaymneh, Khaled and Wu, Xiaohua and Couillard, Martin and
               Korkusinski, Marek and Sch{\"o}ll, Eva and J{\"o}ns, Klaus D and
               Zwiller, Valery and Williams, Robin L",
  journal   = "Nano Lett.",
  publisher = "American Chemical Society (ACS)",
  volume    =  18,
  number    =  5,
  pages     = "3047--3052",
  month     =  may,
  year      =  2018,
  language  = "en"
}

@ARTICLE{Wyborski2023-rm,
  title     = "Impact of {MBE-grown} ({In,Ga)As/GaAs} metamorphic buffers on
               excitonic and optical properties of single quantum dots with
               single-photon emission tuned to the telecom range",
  author    = "Wyborski, Pawe{\l} and Gawe{\l}czyk, Micha{\l} and Podemski,
               Pawe{\l} and Wro{\'n}ski, Piotr Andrzej and Pawlyta,
               Miros{\l}awa and Gorantla, Sandeep and Jabeen, Fauzia and
               H{\"o}fling, Sven and S{\k e}k, Grzegorz",
  journal   = "Phys. Rev. Appl.",
  publisher = "American Physical Society (APS)",
  volume    =  20,
  number    =  4,
  month     =  oct,
  year      =  2023,
  copyright = "https://link.aps.org/licenses/aps-default-license",
  language  = "en"
}

@article{Srocka2020,
  title = {Deterministically fabricated quantum dot single-photon source emitting indistinguishable photons in the telecom {O}-band},
  volume = {116},
  ISSN = {1077-3118},
  url = {http://dx.doi.org/10.1063/5.0010436},
  DOI = {10.1063/5.0010436},
  number = {23},
  journal = {Applied Physics Letters},
  publisher = {AIP Publishing},
  author = {Srocka,  N. and Mrowiński,  P. and Große,  J. and von Helversen,  M. and Heindel,  T. and Rodt,  S. and Reitzenstein,  S.},
  year = {2020},
  month = jun 
}

@misc{Gaur2025,
  doi = {10.48550/ARXIV.2512.12300},
  url = {https://arxiv.org/abs/2512.12300},
  author = {Gaur,  Kartik and Barua,  Avijit and Tripathi,  Sarthak and Roche,  Léo J. and Wilksen,  Steffen and Steinhoff,  Alexander and Baraz,  Sam and Nitin,  Neha and Palekar,  Chirag C. and Koulas-Simos,  Aris and Limame,  Imad and Mudi,  Priyabrata and Rodt,  Sven and Gies,  Christopher and Reitzenstein,  Stephan},
  title = {Scalable Quantum Photonic Platform Based on Site-Controlled Quantum Dots Coupled to Circular Bragg Grating Resonators},
  publisher = {arXiv},
  year = {2025},
  copyright = {Creative Commons Attribution 4.0 International}
}

@article{Choquette1997,
  title = {Advances in selective wet oxidation of {AlGaAs} alloys},
  volume = {3},
  ISSN = {1077-260X},
  url = {http://dx.doi.org/10.1109/2944.640645},
  DOI = {10.1109/2944.640645},
  number = {3},
  journal = {IEEE Journal of Selected Topics in Quantum Electronics},
  publisher = {Institute of Electrical and Electronics Engineers (IEEE)},
  author = {Choquette,  K.D. and Geib,  K.M. and Ashby,  C.I.H. and Twesten,  R.D. and Blum,  O. and Hou,  H.Q. and Follstaedt,  D.M. and Hammons,  B.E. and Mathes,  D. and Hull,  R.},
  year = {1997},
  month = jun,
  pages = {916–926}
}

@article{Albert2010,
  title = {Quantum efficiency and oscillator strength of site-controlled {InAs} quantum dots},
  volume = {96},
  ISSN = {1077-3118},
  url = {http://dx.doi.org/10.1063/1.3393988},
  DOI = {10.1063/1.3393988},
  number = {15},
  journal = {Applied Physics Letters},
  publisher = {AIP Publishing},
  author = {Albert,  F. and Stobbe,  S. and Schneider,  C. and Heindel,  T. and Reitzenstein,  S. and H\"{o}fling,  S. and Lodahl,  P. and Worschech,  L. and Forchel,  A.},
  year = {2010},
  month = apr 
}

@article{Mntynen2019,
  title = {Single-photon sources with quantum dots in {III–V} nanowires},
  volume = {8},
  ISSN = {2192-8606},
  url = {http://dx.doi.org/10.1515/nanoph-2019-0007},
  DOI = {10.1515/nanoph-2019-0007},
  number = {5},
  journal = {Nanophotonics},
  publisher = {Walter de Gruyter GmbH},
  author = {M\"{a}ntynen,  Henrik and Anttu,  Nicklas and Sun,  Zhipei and Lipsanen,  Harri},
  year = {2019},
  month = apr,
  pages = {747–769}
}

@article{Kim2016,
  title = {Two-photon interference from a bright single-photon source at telecom wavelengths},
  volume = {3},
  ISSN = {2334-2536},
  url = {http://dx.doi.org/10.1364/OPTICA.3.000577},
  DOI = {10.1364/optica.3.000577},
  number = {6},
  journal = {Optica},
  publisher = {Optica Publishing Group},
  author = {Kim,  Je-Hyung and Cai,  Tao and Richardson,  Christopher J. K. and Leavitt,  Richard P. and Waks,  Edo},
  year = {2016},
  month = jun,
  pages = {577}
}

@article{Podhorsk2026,
  title = {Buried Stressor Engineering for Position-Controlled {InGaAs} Quantum Dots with Local Density Variation for Integrated Quantum Photonics},
  volume = {13},
  ISSN = {2330-4022},
  url = {http://dx.doi.org/10.1021/acsphotonics.5c02303},
  DOI = {10.1021/acsphotonics.5c02303},
  number = {2},
  journal = {ACS Photonics},
  publisher = {American Chemical Society (ACS)},
  author = {Podhorský,  Martin and Klonz,  Maximilian and B\"{o}hmer,  Lux and Kulig,  Sebastian and Palekar,  Chirag C. and Klenovský,  Petr and Rodt,  Sven and Reitzenstein,  Stephan},
  year = {2026},
  month = jan,
  pages = {471–481}
}

@article{Klenovsk2026,
  title = {Coulomb correlated multiparticle states of weakly confining {GaAs} quantum dots},
  volume = {113},
  ISSN = {2469-9969},
  url = {http://dx.doi.org/10.1103/wjw9-blb1},
  DOI = {10.1103/wjw9-blb1},
  number = {12},
  pages = {125410},
  journal = {Physical Review B},
  publisher = {American Physical Society (APS)},
  author = {Klenovský,  Petr},
  year = {2026},
  month = mar 
}

@article{Schweickert2018,
  title = {On-demand generation of background-free single photons from a solid-state source},
  volume = {112},
  ISSN = {1077-3118},
  url = {http://dx.doi.org/10.1063/1.5020038},
  DOI = {10.1063/1.5020038},
  number = {9},
  journal = {Applied Physics Letters},
  publisher = {AIP Publishing},
  author = {Schweickert,  Lucas and J\"{o}ns,  Klaus D. and Zeuner,  Katharina D. and Covre da Silva,  Saimon Filipe and Huang,  Huiying and Lettner,  Thomas and Reindl,  Marcus and Zichi,  Julien and Trotta,  Rinaldo and Rastelli,  Armando and Zwiller,  Val},
  year = {2018},
  month = feb 
}

@article{Semenova2008,
  title = {Metamorphic approach to single quantum dot emission at 1.55\textmu m on {GaAs} substrate},
  volume = {103},
  ISSN = {1089-7550},
  url = {http://dx.doi.org/10.1063/1.2927496},
  DOI = {10.1063/1.2927496},
  number = {10},
  journal = {Journal of Applied Physics},
  publisher = {AIP Publishing},
  author = {Semenova,  E. S. and Hostein,  R. and Patriarche,  G. and Mauguin,  O. and Largeau,  L. and Robert-Philip,  I. and Beveratos,  A. and Lemaître,  A.},
  year = {2008},
  month = may 
}

@article{Senellart2017,
  title = {High-performance semiconductor quantum-dot single-photon sources},
  volume = {12},
  ISSN = {1748-3395},
  url = {http://dx.doi.org/10.1038/nnano.2017.218},
  DOI = {10.1038/nnano.2017.218},
  number = {11},
  journal = {Nature Nanotechnology},
  publisher = {Springer Science and Business Media LLC},
  author = {Senellart,  Pascale and Solomon,  Glenn and White,  Andrew},
  year = {2017},
  month = nov,
  pages = {1026–1039}
}

@article{Ding2016,
  title = {On-Demand Single Photons with High Extraction Efficiency and Near-Unity Indistinguishability from a Resonantly Driven Quantum Dot in a Micropillar},
  volume = {116},
  ISSN = {1079-7114},
  url = {http://dx.doi.org/10.1103/PhysRevLett.116.020401},
  DOI = {10.1103/physrevlett.116.020401},
  number = {2},
  journal = {Physical Review Letters},
  publisher = {American Physical Society (APS)},
  author = {Ding,  Xing and He,  Yu and Duan,  Z.-C. and Gregersen,  Niels and Chen,  M.-C. and Unsleber,  S. and Maier,  S. and Schneider,  Christian and Kamp,  Martin and H\"{o}fling,  Sven and Lu,  Chao-Yang and Pan,  Jian-Wei},
  year = {2016},
  month = jan 
}

@article{Sapienza2015,
  title = {Nanoscale optical positioning of single quantum dots for bright and pure single-photon emission},
  volume = {6},
  ISSN = {2041-1723},
  url = {http://dx.doi.org/10.1038/ncomms8833},
  DOI = {10.1038/ncomms8833},
  number = {1},
  journal = {Nature Communications},
  publisher = {Springer Science and Business Media LLC},
  author = {Sapienza,  Luca and Davan\c{c}o,  Marcelo and Badolato,  Antonio and Srinivasan,  Kartik},
  year = {2015},
  month = jul 
}

@article{Mller2014,
  title = {On-demand generation of indistinguishable polarization-entangled photon pairs},
  volume = {8},
  ISSN = {1749-4893},
  url = {http://dx.doi.org/10.1038/nphoton.2013.377},
  DOI = {10.1038/nphoton.2013.377},
  number = {3},
  journal = {Nature Photonics},
  publisher = {Springer Science and Business Media LLC},
  author = {M\"{u}ller,  M. and Bounouar,  S. and J\"{o}ns,  K. D. and Gl\"{a}ssl,  M. and Michler,  P.},
  year = {2014},
  month = feb,
  pages = {224–228}
}
\addcontentsline{toc}{section}{References}
\clearpage

\end{document}